\documentclass[12pt,a4paper,final]{iopart}

\usepackage{iopams}  
\usepackage{graphicx}
\usepackage[dvipdfm,breaklinks=true,colorlinks=true,linkcolor=blue,urlcolor=blue,citecolor=blue]{hyperref}
\usepackage{cite}

\begin{document}

\title[Gravity in Extreme Regions Based on NC Quantization of TG]{Gravity in Extreme Regions Based on Noncommutative Quantization of Teleparallel Gravity}

\author{Ryouta Matsuyama$^1$ and Michiyasu Nagasawa$^{1,2}$}
\address{$^1$Department of Physics, Graduate School of Science, Kanagawa University, Kanagawa 259-1293, Japan}
\address{$^2$Department of Physics, Faculty of Science, Kanagawa University, Kanagawa 259-1293, Japan}
\ead{r201770192ve@jindai.jp}



\begin{abstract}
In this paper, a noncommutative gravitational theory is constructed by applying Moyal deformation quantization and the Seiberg--Witten map to teleparallel gravity,  a classical gravitational theory, as a gauge theory of local translational symmetry.
Since our model is based on teleparallel gravity, it is an extremely simple noncommutative gravitational theory.
We also clearly divide the role of the products, such that the metric is responsible for the rule of the inner product (which is calculated by taking the sum over the subscripts) and the Moyal product is responsible for tensor and field noncommutativity.
This solves problems related to the order of the products and the relationship between the metric and the Moyal product.
Furthermore, we analyze the cosmic evolution of the very early universe and the spacetime features around black holes using the constructed noncommutative gravitational theory, and find that gravity acts repulsively in the extreme region where its quantum effects become prominent.

\end{abstract}
\vspace{2pc}

\submitto{\CQG}

\section{Introduction}
The establishment of a quantum gravitational theory and a unified field theory including gravity is an imperative problem in modern physics.
Such theories describe gravitational interactions in extreme cases, such as the earliest stage of the universe or the vicinity of the spacetime singularity.
Many physicists have been exploring such theories, but definitive candidates are yet to be discovered.

In quantum theory, physical quantities are discretized; one reason for this is that they become noncommutative.
Position and momentum are defined as operators in quantum theory and have been shown to be noncommutative; their commutation relations are defined.
As a consequence, the characteristics of quantum theory, such as the uncertainty principle and discretization of physical quantities, are derived.
Snyder was the first to present a spacetime-quantization method focusing on such noncommutativity and commutation relations \cite{1}.
By defining the position operator based on the assumption that the momentum space is de Sitter, he led the spacetime coordinates become noncommutative while maintaining Lorentz invariance.
In fact, his noncommutative spacetime was originally developed as a method to solve ultraviolet divergence in quantum electrodynamics, but because the method of renormalization had achieved great success, his approach was almost neglected.
However, in the superstring theory, which rapidly evolved in the 1990s, it was shown that noncommutative spacetime exists in a very natural form \cite{6}.
Also in the mathematics field, Connes showed that noncommutative geometry can form the basis of quantum theory at the same time \cite{2,30}.
As a result, noncommutative spacetime research has become active in various fields, including mathematics and physics.
Exploration of gravitational theory based on noncommutative spacetime has become a major topic in physics and is the subject of this paper.

A general method used to construct gravitational theory based on noncommutative spacetime is the deformation quantization using the Moyal product \cite{36}.
Unlike canonical quantization, which defines the commutation relation by converting the physical quantity into an operator, deformation quantization is a method of redefining the product of functions without such a convention.
Since the Moyal product of functions is noncommutative, it is possible to construct a gravitational theory that includes the nontrivial commutation relation by replacing the ordinary product present in classical gravitational theory with a Moyal product.
In the literature, deformation quantization is used for various forms of gravitational theory, and there are two main streams.
The first is a method using the concept of a twist with a modified Leibniz rule of gauge transformation \cite{9,10} and the second is a method using a Seiberg--Witten map which is the transformation formula of the gauge theory on both spacetime \cite{8,40}.
Another study \cite{48} using the Seiberg--Witten map shows the coupling of gravity and fermion on noncommutative spacetime and gives expansion  of noncommutative parameter up to the second order.
Unlike the above methods, one previous study has simply incorporated the Moyal product into complex symplectic gravity \cite{7}.
The important point in these studies using the deformation-quantization method is that they aim to reproduce general relativity on a noncommutative spacetime.

The best known classical gravitational theory is general relativity, but there is another theory, called teleparallel gravity \cite{11,12,51,53}.
Teleparallel gravity is a gauge theory on local translational symmetry, which is different from general relativity treating gravity geometrically \cite{11,60}.
However, the field equation is equal to the Einstein equation, and the Lagrangian of both theories is also equivalent \cite{11,12,61}.
Therefore, it is also called teleparallel equivalent of general relativity.
Note that since tetrad has extra degrees of freedom, teleparallel gravity is quasi-Lorentz invariant \cite{52,57}, so the modified theory such as the $f(T)$ theory is not Lorentz invariant \cite{52,57,58,62}.
Teleparallel gravity is studied vigorously especially in fields related to modified gravitational theory \cite{63,64,65}.
Also, in the modified theory, studies motivated by quantum gravity are also conducted \cite{54,55,56}.

Difficulty in constructing a quantum gravitational theory has been partly attributed to the geometrical properties of general relativity \cite{44,45}, so it can be naively inferred that teleparallel gravity is superior to general relativity as a quantization target.
Also, since teleparallel gravity is a gauge theory, it has an advantageous feature that it is possible to apply the previous research of the gauge theory on flat noncommutative spacetime.
In \cite{38}, the authors clarified that the Weitzenb{\"o}ck space can be derived by dimensional reduction of gauge theory on noncommutative spacetime, thereby showing the compatibility of teleparallel gravity with such a spacetime.
Additionally, in \cite{39}, noncommutative gravitational theory, in which torsion becomes field strength, is constructed on a symplectic manifold in which the covariant star product is defined.
Unlike these studies, \cite{13,14} perturbatively derived  the Lagrangian responsible for noncommutative gravitational effects by applying the deformation quantization and Seiberg--Witten map to teleparallel gravity.
However, these previous studies have some unresolved problems with the order of products, the relationship between the star product and the metric, the validity of the Seiberg--Witten map when noncommutative parameters are covariant, and the definition of Einstein's summation convention.

Therefore, in this paper, we apply Moyal-deformation quantization to teleparallel gravity according to \cite{11} and construct a gravitational theory on the noncommutative spacetime.
The characteristics of this theory are extremely simple because the tetrad and metric are guided naturally by determining the gauge field.
Additionally, we clearly divide the roles of the product such that the metric and tetrad are responsible for the inner product by taking the sum over the subscripts, while the Moyal product expresses the noncommutativity of spacetime, thereby avoiding difficulties with product order.
Finally, by applying the noncommutative gravitational theory constructed in this paper to astrophysics and cosmology, we reveal the tendency of quantum-gravitational effects based on a noncommutative spacetime.

The rest of the paper is organized as follows.
Section \ref{sec2} describes the gauge theory of noncommutative spacetime based on the Moyal product.
Section \ref{sec3} introduces teleparallel gravity, which is a classical gravitational theory.
In the subsequent chapters of this paper, we mention original work.
Section \ref{sec4} explains noncommutative quantization of gravity, which is the main subject of this paper.
Sections \ref{sec5} and \ref{sec6} discuss gravity in extreme regions where quantum effect becomes prominent, in order to investigate the difference between the noncommutative and classical gravitational theories.
In Section \ref{sec5}, we apply noncommutative gravity to cosmology and investigate the effect of spacetime noncommutativity in the earliest stage of the universe.
In Section \ref{sec6}, we refer to the noncommutative quantum effect in the vicinity of the event horizon and discuss its observability.
Finally, we conclude this paper in Section \ref{sec7}.

\section{\label{sec2}Gauge Theory on Noncommutative Spacetime}
In this section, we briefly review gauge theory and the Seiberg--Witten map on a noncommutative spacetime, as described in \cite{16,41,19,21,42,43,37}.
Note that unless otherwise mentioned,``noncommutativity" in this paper is a concept distinct from gauge-group noncommutativity, which refers to gauge groups that are non-Abelian.

A noncommutative spacetime satisfies the nontrivial commutation relation with coordinates. 
In this paper, deformation quantization by the Moyal product is used to construct the noncommutative spacetime.
The product of arbitrary functions $f(x), g(x) $ using the Moyal product $\star$ is defined as 
\begin{eqnarray}
\nonumber
( f \star g )(x) &=& \exp \bigg[ {\frac{i}{2}\theta^{\mu\nu} \frac{\partial}{\partial \alpha^\mu} \frac{\partial}{\partial \beta^\nu}}  \bigg] f(x+ \alpha)\ g(x + \beta ) \bigg|_{\alpha=\beta} \\
\label{2.12}
&=& fg + \frac{i}{2}\theta^{\mu\nu}\partial_\mu f \ \partial_\nu g + {\cal O}(\theta^2) ,
\end{eqnarray}
 whereby the spacetime coordinate $x^\mu (\mu = 0,1,2,3)$ satisfies the nontrivial commutation relation 
 \begin{eqnarray}
\label{2.19}
[x^\mu , x^\nu]_\star = x^\mu \star x^\nu - x^\nu \star x^\mu = i\theta^{\mu\nu},
\end{eqnarray}
where $\theta^{\mu\nu}$ is a quantity commonly referred to the noncommutative parameter and is antisymmetric with respect to the exchange of subscripts, that is $\theta^{\mu\nu} =  -\theta^{\nu\mu}$.
In this paper, all the components of noncommutative parameter is assumed to be a constant, and if the influence of spacetime noncommutativity is equated to the spacetime quantum effect, the square root of all the components of noncommutative parameter is regarded as being on the order of Planck length, $\sqrt{\theta^{\mu\nu}} \sim l_{\rm Pl}$.
Therefore, the noncommutative parameter is sufficiently small, when the system of physical scales much larger than $l_{\rm Pl}$ is considered thus we ignore terms of second and higher order of $\theta^{\mu\nu}$.

The gauge transformation of any gauge field $ \hat{A}_\mu$ on the noncommutative spacetime is expressed as 
\begin{eqnarray}
\label{3.2}
{\delta}_{\hat{\lambda}} \hat{A}_\mu = \partial_\mu \hat{\lambda}+ i[ \hat{\lambda} , \hat{A}_\mu  ]_\star \ ,
\end{eqnarray}
where $\hat \lambda$ is the noncommutative parameter.
Importantly, the commutation relation in ordinary gauge theory is defined only in terms of the gauge group, but on noncommutative spacetime, it is defined based on the gauge group as well as spacetime noncommutativity.
In addition, field strength can be defined as 
\begin{eqnarray}
\label{3.4}
\hat{F}_{\mu\nu} \equiv i [\hat{D}_\mu ,\hat{D}_\nu]_\star = \partial_\mu \hat{A}_\nu - \partial_\nu \hat{A}_\mu -i [ \hat{A}_\mu \ , \hat{A}_\nu]_\star ,
\end{eqnarray}
using covariant derivative $\hat{D}_\mu \equiv \partial_\mu  - i {\hat A}_\mu$. Its gauge transformation is 
\begin{eqnarray}
\label{3.5}
{\delta}_{\hat{\lambda}} \hat{F}_{\mu\nu} = i[\hat{\lambda}, \hat{F}_{\mu\nu}]_\star  .
\end{eqnarray}
Furthermore, the action, $S$, and the Lagrangian density, ${\cal L}$, in the case of a flat noncommutative spacetime with a Minkowski metric $\eta_{\mu\nu}=(-1,+1,+1,+1)$ are defined as \cite{6,16,41,42,43}
\begin{eqnarray}
\nonumber
S = \int d^4 x {\cal L} &=& \frac{1}{4} \int d^4 x\  \eta^{\mu\nu} \eta^{\rho\sigma} {\hat F}_{\mu\rho} \star {\hat F}_{\nu\sigma} \\
\label{3.6}
    &=& \frac{1}{4} \int d^4 x\   {\hat F}_{\mu\nu} \star {\hat F}^{\mu\nu} \ .
\end{eqnarray}
The invariance of action on gauge transformation is satisfied, if cyclic symmetry \cite{6,16,41,42,43}
\begin{eqnarray}
\label{3.21}
\int d^4x\ f \star g \star h = \int d^4x\ g \star h \star f = \int d^4x\ h \star f \star g,
\end{eqnarray}
holds.

Although the gauge theory on the noncommutative spacetime shown here seems to be the same as that of the commutative spacetime, its mathematical structure has been significantly changed.
On the commutative spacetime, the generator $T_A$ of gauge symmetry consists of the elements of Lie algebra ${\cal G}$, whereas on the noncommutative spacetime, it is extended to the enveloping algebra, ${\cal U(G)}$, of  ${\cal G}$.
In this paper, subscripts with uppercase alphabets $(A, B, \ldots)$ are used when they represent internal space.
First, since the generator on commutative spacetime is $ T_A \in {\cal G} $, these commutators are 
\begin{eqnarray}
\label{3.10}
[T_A , T_B] = T_A T_B - T_B T_A .
\end{eqnarray}
Also, because the gauge parameters and gauge fields can be expanded in terms of $T_A$  as $\lambda = \lambda^A T_A$ or $A_\mu = A^A{_\mu}T_A$, these commutators can be written as 
\begin{eqnarray}
\label{3.11}
i [ \lambda , A_\mu ] = i\lambda^A A^B{_\mu}[T_A, T_B] 
\end{eqnarray} 
on a commutative spacetime.
On the other hand, in the case of $T_A \in {\cal U} ({\cal G})$, the commutation and anti-commutation relations of the generator are defined using tensor products such as 
\begin{eqnarray}
\label{3.14}
[T_A, T_B] &=& T_A \otimes T_B - T_B \otimes T_A  ,\\
\label{3.22}
  \{T_A, T_B\}  &=& T_A \otimes T_B + T_B \otimes T_A .
\end{eqnarray}
On noncommutative spacetime, the gauge parameters and gauge fields are ordered by Moyal product, in contrast to commutative spacetime.
Therefore, in order to assign generators corresponding to gauge parameters and gauge fields, the Lie algebra must be extended to the enveloping algebra for which the tensor product is defined.
As a result, the commutation relation between the gauge parameter, $\hat{\lambda} = \hat{\lambda}^A T_A$, and the gauge field, $\hat{A}_\mu = \hat{A}^A{_\mu}T_A$, on the noncommutative spacetime can be expressed as 
\begin{eqnarray}
\label{3.19}
i [ \hat{\lambda}, \hat{A}_\mu]_\star = \frac{i}{2} [ \hat{\lambda}^A, \hat{A}^B{_\mu}]_\star \{T_A,T_B\} + \frac{i}{2} \{ \hat{\lambda}^A, \hat{A}^B{_\mu} \}_\star [T_A,T_B].
\end{eqnarray}
From (\ref{3.19}), it is absolutely necessary to define not only the commutation relation of the generator but also the anti-commutation relation on the noncommutative spacetime.
Therefore, in this paper, the commutation and anti-commutation relations of the generators are defined as 
\begin{eqnarray}
\label{3.20}
[T_A,T_B] = i f^C{_{AB}} T_C , \hspace{10mm} \{T_A,T_B\} =  g^C{_{AB}} T_C,
\end{eqnarray}
according to \cite{13,14,40,70}, where $f^C{_{AB}}$ and $g^C{_{AB}} $ are the structure constants of the commutator and anti-commutator, respectively.

\subsection{\label{subsec1}Seiberg--Witten Map}
There is a correspondence between the gauge theories on commutative and noncommutative spacetimes \cite{6}.
In particular, from the definition of the Moyal product (\ref{2.12}), it is possible to equate gauge theories on a noncommutative spacetime and on an ordinary spacetime where  $\theta^{\mu\nu}$ exists.
In other words, there is a map, known as the Seiberg--Witten map, connecting these theories, which are considered equivalent.
Thus, it is possible to handle noncommutative-spacetime physics on an ordinary spacetime.
We summarize this map for an Abelian gauge theory in order to apply it to teleparallel gravity.

According to the Seiberg--Witten map, since the gauge parameters, $\hat{\lambda}$, and gauge fields, $\hat{A}_\mu$, on the noncommutative spacetime are functions of classical gauge parameter $\lambda$ and gauge field $A_\mu$, they are defined as 
\begin{eqnarray}
\label{4.2}
\hat{\lambda}(A,\lambda) &=& \lambda - i \theta^{\mu\nu} \lambda^{(1)}_{\mu\nu}(A,\lambda), \\
\label{4.1}
\hat{A}_\rho (A) &=& A_\rho - i \theta^{\mu\nu}A^{(1)}_{\rho\mu\nu}(A),
\end{eqnarray}
where $\lambda^{(1)}_{\mu\nu}$ and $A^{(1)}_{\rho\mu\nu}$ are the first order term.
In addition, the relation between the gauge fields of two spacetimes is represented as
\begin{eqnarray}
\label{4.12}
\hat{A}(A+\delta_\lambda A) = \hat{A}(A) + {\delta}_{\hat{\lambda}} \hat{A}(A) ,
\end{eqnarray}
and the gauge parameter on the noncommutative spacetime satisfies 
\begin{eqnarray}
\label{4.5}
i \hat{\lambda}(i [ {\alpha} , {\beta} ]) = i\delta_{{\alpha}} \hat{\lambda}(\beta)  - i\delta_{{\beta}} \hat{\lambda}(\alpha) + [ \hat{\lambda}(\alpha) , \hat{\lambda}(\beta) ]_\star ,
\end{eqnarray}
where $\delta_\lambda {\hat \Phi} = \delta_{\hat \lambda} {\hat \Phi}$.
By substituting (\ref{4.2}) and (\ref{4.1}) into (\ref{4.12}) and (\ref{4.5}), ${\hat \lambda}$ and ${\hat A}_\rho$ are derived as \cite{42,43,37}:
\begin{eqnarray}
\label{4.22}
\hat{\lambda} &=& \lambda + \frac{1}{2} \theta^{\mu\nu}  \partial_\mu \lambda \ A_\nu,   \\
\label{4.23}
\hat{A}_\rho &=& A_\rho -\frac{1}{2} \theta^{\mu\nu}  A_\mu ( \partial_\nu A_\rho + F_{\nu\rho} ),
\end{eqnarray}
where $F_{\nu\rho} = \partial_\nu A_\rho - \partial_\rho A_\nu$ is classical field strength.
Since ${\hat A}_\rho$ is expressed as a function of $ A_\rho$, the field strength $\hat{F}_{\mu\nu}$ on the noncommutative spacetime can be expressed as 
\begin{eqnarray}
\label{4.26}
\hat{F}_{\rho\sigma} = F_{\rho\sigma} 
          + \theta^{\mu\nu} \big( F_{\rho\mu}F_{\sigma\nu} - A_\mu \partial_\nu F_{\rho\sigma} \big).
\end{eqnarray}
At this time, since the classical gauge transformation according to $\delta_\lambda A_\rho = \partial_\rho \lambda$ of (\ref{4.26}) is expressed as 
\begin{eqnarray}
\label{4.27}
\delta_{{\lambda}} \hat{F}_{\rho\sigma} = - \theta^{\mu\nu} \partial_\mu \lambda \partial_\nu F_{\rho\sigma},
\end{eqnarray}
it is obvious that $\delta_{\hat \lambda} \hat{F}_{\rho\sigma} = \delta_{\lambda} \hat{F}_{\rho\sigma}$ is satisfied by comparison with (\ref{3.5}).
By applying the above technique to the matter field $\hat{\Psi}$ on the noncommutative spacetime, we can derive its Seiberg--Witten map.
Like ${\hat A}_\rho$ and ${\hat \lambda}$, $\hat{\Psi}$ can be expanded as
\begin{eqnarray}
\label{4.19}
\hat{\Psi}(\Psi,A) = \Psi  - i \theta^{\mu\nu} \Psi^{(1)}_{\mu\nu}(\Psi, A),
\end{eqnarray}
where $\Psi^{(1)}_{\mu\nu}$ is the first order term of matter field.
Also, since the gauge transformation of $\hat{\Psi}$ is
\begin{eqnarray}
\label{4.25}
\delta_{\hat{\lambda}} \hat{\Psi} = i \hat{\lambda} \star \hat{\Psi},
\end{eqnarray}
by substituting (\ref{4.19}) into (\ref{4.25}), the Seiberg--Witten map of $\hat \Psi$ can be derived as
\begin{eqnarray}
\label{4.24}
\hat{\Psi} = \Psi - \frac{1}{2}\theta^{\mu\nu} A_\mu \partial_\nu \Psi .
\end{eqnarray}

\section{\label{sec3}Teleparallel Gravity}
In this section, we describe teleparallel gravity, a classical gravitational theory that is the basis of noncommutative quantization using the Seiberg--Witten map of \cite{11}.
Teleparallel gravity is an alternative theory of gravity and it is also known as teleparallel equivalent of general relativity because its field equation is equivalent to the Einstein equation.
However, since teleparallel gravity is a gauge theory constructed with local translation gauge symmetry, it is conceptually different from general relativity in which gravity is interpreted geometrically \cite{11,60}.
Note that it is impossible to find a modified general relativity equivalent to the modified teleparallel gravity with a naive method of showing the equivalence between teleparallel gravity and general relativity \cite{52,57,58,62}.

Gauge transformation is defined on the internal space, and the local translational transformation is related to the coordinate system on actual spacetime.
These are seemingly different transformations.
Therefore, teleparallel gravity defines tangent space at each point in spacetime corresponding to the internal space of gauge theory.
Hereafter, $x^\mu$ denotes the spacetime coordinate, $x^a$ is the tangent-space coordinate, $g_{\mu\nu}$ is the metric of the spacetime.
Moreover, the metric of the tangent space is the Minkowski metric, $\eta_{ab}={\rm diag}(-1,+1,+1,+1)$.
Hereafter, unless otherwise noted, lower case alphabets $(a, b, \ldots)$ are used when representing tangent space.
The arbitrary matter field $\Phi(x)$ undergoes the local translational transformation $x^a \rightarrow x'^a = x^a - \epsilon^a$ as
\begin{eqnarray}
\label{5.1}
\Phi(x) \rightarrow \Phi '(x') = \Phi(x) - \epsilon^a \partial_a \Phi(x),
\end{eqnarray}
where $\epsilon^a$ is a parameter of the local translational transformation that depends only on spacetime coordinates.
On the other hand, the general gauge transformation is 
\begin{eqnarray}
\label{5.2}
\Phi(x) \rightarrow \Phi '(x') &= e^{i\lambda}\Phi(x) =\Phi(x) + i\lambda^A T_A\Phi(x) ,
\end{eqnarray}
where $\lambda^A$ and $T_A$ are gauge parameters and generators, respectively.
By comparing (\ref{5.1}) and (\ref{5.2}), it is possible to regard the tangent space as an internal space such that the local translational transformation is identified as a gauge transformation and to define the generator of the local translational symmetry as $P_a = i \partial_a$.
At this time, the gauge parameter of local translational symmetry is $\epsilon = \epsilon^a P_a$.
Here the covariant derivative is defined as
\begin{eqnarray}
h_\mu \equiv \partial _\mu - i B_\mu = \partial_\mu + B^a{_{\mu}} \partial_a,
\end{eqnarray}
where $B^a{_\mu}$ is a gauge field of the local translational transformation. 
Since $B^a{_\mu}$ depends only on spacetime coordinates, it satisfies $[\epsilon, B_\mu] = 0$, so its transformation is 
\begin{eqnarray}
\label{5.5}
\delta_\epsilon B^a{_\mu} = \partial_\mu \epsilon^a .
\end{eqnarray}
In this sense, it is clear that teleparallel gravity behaves as an Abelian gauge theory.
When the covariant differential operator is applied to the coordinates on the tangent space, it becomes
\begin{eqnarray}
\label{5.6}
h_\mu x^a = \partial_\mu x^a + B^a{_\mu}.
\end{eqnarray}
The first term on the right-hand side of (\ref{5.6}) is the matrix of the coordinate transformation.
Since this transformation matrix has subscripts for both spacetime and tangent space, it is responsible for the transformation connecting the two spaces.
In other words, $h_\mu x^a$ connects an arbitrary spacetime on which a gauge field exists and a tangent space.
Therefore, tetrad is defined as 
\begin{eqnarray}
\label{5.7}
h^a{_\mu} & \equiv h_\mu x^a  =e^a{_\mu} + B^a{_\mu},
\end{eqnarray}
where $e^a{_\mu} \equiv  \partial_\mu x^a$ is the trivial part of the tetrad.
Note that $e^a{_\mu}$ is exactly different from Kronecker delta because the coordinate system on spacetime is arbitrary.
Furthermore, since $e^a{_\mu}$ satisfies $\delta_\epsilon e^a{_\mu} = - \partial_\mu \epsilon^a = - \delta_\epsilon B^a{_\mu}$ under the local translation $x^a \rightarrow x^a - \epsilon^a$, the transformation of $h^a{_\mu}$ is 
\begin{eqnarray}
\delta_\epsilon h^a{_\mu} = 0.
\end{eqnarray}
The tetrad notation, $e^a{_{\mu}} = \delta^a_\mu + B^a{_{\mu}}$ (where $e^a{_{\mu}}$ is the nontrivial tetrad and $\delta^a_{\mu}$ is the trivial part of the tetrad) is widely used, but in this paper we instead use (\ref{5.7}), following \cite{11}.
Note that the tetrad and $B^a{_{\mu}}$ are not independent variables; thus, the number of degrees of freedom is the same as in other notational schemes \cite{49,50}.

The connection between the spacetime and the tangent space can be understood by the relation between the tetrad and metric, namely
\begin{eqnarray}
\label{5.8}
g_{\mu\nu} = \eta_{ab} h^a{_\mu} h^b{_\nu}, \hspace{10mm} g^{\mu\nu} = \eta^{ab} h_a{^\mu} h_b{^\nu}.
\end{eqnarray}
Also, $h^a{_\mu}$ and $h_a{^\mu}$ satisfy the orthonormal condition
\begin{eqnarray}
h^a{_\mu} h_a{^\nu} = \delta^\mu_\nu, \hspace{16mm} h^a{_\mu} h_b{^\mu} = \delta^a_b, 
\end{eqnarray}
and the inverse transform of (\ref{5.8})
\begin{eqnarray}
\label{5.10}
\eta_{ab} = g_{\mu\nu}h_a{^\mu}h_b{^\nu} , \hspace{10mm} \eta^{ab} = g^{\mu\nu}h^a{_\mu}h^b{_\nu}.
\end{eqnarray}
With the tetrads defined, it becomes possible to take the inner product between tensors on different spaces, as well as the inner products of tensors in the same space defined by $g_{\mu\nu}$ and $\eta_{ab}$.
In other words, the inner product of the tensors $u = u^\mu \partial_\mu$ and $v=v^\mu \partial_\mu$ on the spacetime is defined by the metric as 
\begin{eqnarray}
\label{5.11}
g( u , v )  =  g ( \partial_\mu , \partial_\nu  )( u^\mu  \otimes v^\nu ) = g_{\mu\nu} ( u^\mu  \otimes v^\nu  ) = g_{\mu\nu}u^\mu v^\nu,
\end{eqnarray}
while the inner product of the tensor $w$ on the tangent space and the tensor $u$ on the spacetime is defined by the tetrad as 
\begin{eqnarray}
\label{5.12}
g( u , w ) = g ( \partial_\mu , dx^a  )( u^\mu  \otimes w_a  ) = h^a{_\mu} ( u^\mu  \otimes w_a  ) = h^a{_\mu} u^\mu  w_a.
\end{eqnarray}

Now that the covariant derivative and the gauge field with the local translational symmetry have been defined, we use these formulae to define the field strength and Lagrangian.
Since teleparallel gravity is an Abelian gauge theory, field strength is written as
\begin{eqnarray}
\label{5.15}
F{_{\mu\nu}} =  F^a{_{\mu\nu}} P_a 
=  (\partial_\mu B^a{_\nu} - \partial_\nu B^a{_\mu}) P_a =  (\partial_\mu h^a{_\nu} - \partial_\nu h^a{_\mu}) P_a.
\end{eqnarray}
In addition, field strength can be expressed as a torsion on the Weitzenb\"ock spacetime such as
\begin{eqnarray}
\label{5.18}
T^\rho{_{\mu\nu}} = h_a{^\rho} (\partial_\mu h^a{_\nu} - \partial_\nu h^a{_\mu}) = h_a{^\rho}F^a{_{\mu\nu}}.
\end{eqnarray}
When defining the Lagrangian, it is necessary to take the sum over the spacetime and tangent-space subscripts using the tetrad.
In Yang-Mills gauge theory, since the spacetime and internal space are independent, the Lagrangian has only one term, namely
\begin{eqnarray}
\label{5.16}
{\cal L}_{gauge} = \frac{1}{4}\eta^{\mu\rho}\eta^{\nu\sigma}F^A{_{\mu\nu}}{F_{A}}_{\rho\sigma} = \frac{1}{4}F^A{_{\mu\nu}}F_A{^{\mu\nu}}.
\end{eqnarray}
However, since teleparallel gravity defines a tetrad, it is necessary to take all possible sums over subscripts of the field strength.
Therefore, the Lagrangian is defined as
\begin{eqnarray}
\nonumber
{\cal L}_{TG}  
&=&  \frac{h}{2\kappa}\bigg( \frac{1}{4} \eta_{ab}g^{\mu\rho}g^{\nu\sigma} +\frac{1}{2} h_a{^\sigma}h_b{^\nu}g^{\mu\rho} - h_a{^\nu}h_b{^\sigma}g^{\mu\rho} \bigg) F^a{_{\mu\nu}} F^b{_{\rho\sigma}} \\
\label{5.17}
&=& \frac{h}{2\kappa}\bigg( \frac{1}{4} T^\rho{_{\mu\nu}} T_\rho{^{\mu\nu}} + \frac{1}{2} T^\rho{_{\mu\nu}} T{^{\nu\mu}}{_\rho} - T^\rho{_{\mu\rho}}T{^{\nu\mu}}{_\nu} \bigg),
\end{eqnarray}
where $h = {\rm det}\ ( h^a{_\mu} )$ is a volume element, and in this paper, we use geometrical units with $ c = 1$, $G = 1 $, such that the coupling constant is $ \kappa = 8 \pi $. 
Moreover, the coefficients of each term are determined by the invariance of the infinitesimal transformation $ h^a{_\mu} \rightarrow h^a{_\mu} + \omega^a{_b}h^b{_\mu}$ (where $\omega^a{_b} = - \omega_b{^a}$) of the Lagrangian \cite{11,12}.
However, a surface term that appear in this transformation are ignored.
According to (\ref{5.18}), Lagrangian in the teleparallel gravity can be seen to have the form of the square of torsion.
At this time, ${\cal L}_{TG}$ and Lagrangian of general relativity are equivalent except for surface term\cite{11}.

\section{\label{sec4}Noncommutative Gravity}
In Section \ref{sec2}, we explained the gauge theory and the Seiberg--Witten map on a noncommutative spacetime, and in the previous section, we showed that classical gravitational theory can be expressed as a pure gauge theory.
By integrating these facts, we can construct a gravitational theory on a noncommutative spacetime.
However, teleparallel gravity deals with a nontrivial spacetime on which there is a gauge field of local translational symmetry, which is an obstacle for noncommutative quantization of gravitational theory.

First, we define the role of the metric in the product.
On a commutative spacetime, the product of the functions can be written as
\begin{eqnarray}
\label{q1}
M(f \otimes g) = f g ,
\end{eqnarray}
where $M$ is the multiplication map.
The inner product of vectors in teleparallel gravity is given as (\ref{5.11}), but when using a map like (\ref{q1}), it can be represented as
\begin{eqnarray}
\label{7.1}
 M( g(u , v ) ) = g_{\mu\nu}  M( u^\mu \otimes v^\nu ) = g_{\mu\nu} u^\mu v^\nu .
\end{eqnarray}
Note here that the inner product taking the sum over the subscripts is defined by the metric, whereas the commutativity of vectors is defined by $M$.
In other words, the metric only serves to take the sum over the subscripts, whereas $M$ decides whether the spacetime is commutative or noncommutative.
In the case of a noncommutative spacetime, the Moyal product can be represented as 
\begin{eqnarray}
\label{7.2}
\hat{M}( f \otimes g ) = f \star g ,
\end{eqnarray} 
where $\hat M$ is the Moyal multiplication map.
Therefore, the inner product of vectors $\hat u$ and $\hat v$ on the noncommutative spacetime can be defined as 
\begin{eqnarray}
\label{7.3}
{\hat M} ( {\hat g}( \hat{u} , {\hat v})) \equiv {\hat g}_{\mu\nu} {\hat M} ( {\hat u}^\mu \otimes {\hat v}^\nu ) = {\hat g}_{\mu\nu} ( {\hat u}^\mu \star {\hat v}^\nu ),
\end{eqnarray}
where $ {\hat g}_{\mu\nu}$ is the metric on the noncommutative spacetime.
A similar method as in (\ref{7.3}) is used for defining twisted Lorentz symmetry on a flat noncommutative spacetime when  ${\hat g}_{\mu\nu} = \eta_{\mu\nu}$ \cite{22}.
In this paper, by extending this method to the nontrivial spacetime case, we define the relation between arbitrary metrics on the noncommutative spacetime and the Moyal product.
Note that, there is the relationship between ${\hat g}_{\mu\nu}$ and Moyal product as
\begin{eqnarray}
\label{7.7}
{\hat g}_{\mu\nu} ( {\hat u}^\mu \star {\hat v}^\nu ) \neq  ({\hat g}_{\mu\nu}  {\hat u}^\mu) \star {\hat v}^\nu \neq  {\hat u}^\mu \star ( {\hat g}_{\mu\nu}  {\hat v}^\nu ).
\end{eqnarray}
In addition, the inner-product rule is applied to tetrad ${\hat h}^a{_\mu} $ on a noncommutative spacetime as
\begin{eqnarray}
\label{7.4}
{\hat M} ( {\hat g}( \hat{u} , {\hat s})) = {\hat h}^a{_\mu} {\hat M} ({\hat u}^\mu \otimes {\hat s}_a) =  {\hat h}^a{_\mu}  ({\hat u}^\mu \star {\hat s}_a),
\end{eqnarray}
where ${\hat s}$ is a vector on the tangent space.
Also there exists the relation between ${\hat h}^a{_\mu}$ and Moyal product as
\begin{eqnarray}
\label{7.8}
{\hat h}^a{_\mu}  ({\hat u}^\mu \star {\hat s}_a) \neq  ({\hat h}^a{_\mu} {\hat u}^\mu) \star {\hat s}_a \neq    {\hat u}^\mu \star( {\hat h}^a{_\mu} {\hat s}_a),
\end{eqnarray}
similarly to (\ref{7.7}).
The important point in (\ref{7.3}) and (\ref{7.4}) is that ${\hat g}_{\mu\nu}$ and ${\hat h}^a{_\mu}$ are responsible just for the inner product rule.
Therefore, the relation between ${\hat g}_{\mu\nu}$ and ${\hat h}^a{_\mu}$ is defined without the Moyal product, as
\begin{eqnarray}
\label{7.5}
{\hat g}_{\mu\nu} \equiv \eta_{ab}{\hat h}^a{_{\mu}}{\hat h}^b{_{\nu}}, \hspace{5mm}  \eta_{ab} \equiv {\hat g}_{\mu\nu} {\hat h}_a{^{\mu}}{\hat h}_b{^{\nu}} .
\end{eqnarray}
Also, ${\hat h}_b{^{\mu}}$, which is the inverse tensor of ${\hat h}^a{_\mu}$, is defined to satisfy 
\begin{eqnarray}
\label{7.6}
{\hat h}^a{_{\mu}} {\hat h}_b{^{\mu}} \equiv \delta^a_b, \hspace{12mm}  {\hat h}^a{_{\mu}} {\hat h}_a{^{\nu}} \equiv \delta^\nu_\mu .
\end{eqnarray}
In addition, $ {\hat g}_{\mu\nu} $ is symmetrical with respect to the subscript exchange, as evident from (\ref{7.5}).

\subsection{\label{subsec2}Noncommutative Teleparallel Gravity}
In this subsection, we will construct a noncommutative gravitational theory by referring to the gauge theory on the noncommutative spacetime described in Section \ref{sec2} and by noncommutative quantization of teleparallel gravity.
First, the local translational transformation on the noncommutative spacetime can be expressed as 
\begin{eqnarray}
\label{8.1}
\delta_{\hat \epsilon} {\hat \Phi} = i {\hat \epsilon} \star {\hat \Phi} = - {\hat \epsilon}^a \star \partial_a {\hat \Phi}.
\end{eqnarray}
Also, the transformation of the noncommutative gauge field $ {\hat B}{_\mu} = {\hat B}^a{_\mu} P_a$ is
\begin{eqnarray}
\label{8.2}
\delta_{\hat \epsilon} {\hat B}{_\mu} = \partial_\mu {\hat \epsilon} + i [{\hat \epsilon} , {\hat B}_\mu]_\star 
= \bigg( \partial_\mu {\hat \epsilon}^a + \frac{i}{2} [{\hat \epsilon}^b , {\hat B}^c{_\mu}]_\star g^a{_{bc}} \bigg) P_a.
\end{eqnarray}
As mentioned in Section \ref{sec2}, although classical teleparallel gravity is an Abelian gauge theory, the second term on the right-hand side appears due to the influence of the Moyal product.
Therefore, the relation as (\ref{3.20}) is necessary to construct gauge theory on noncommutative spacetime.
In the literature \cite{13,14,40,70}, the structure constants of anti-commutation relation are given, particularly in \cite{13,14}, based on the generator of teleparallel gravity.
Similarly to preceding works, we define the commutation and anti-commutation relationship of translational generators as
\begin{eqnarray}
\label{8.8}
[P_b, P_c] = 0, \hspace{10mm} \{ P_b, P_c\} = g^a{_{bc}}P_a,
\end{eqnarray}
where $g{_{abc}}$ is invariant under any subscript exchange.
At present, the value of each components of the structure constant of anti-commutation relation is unestablished, and the model of the foregoing study has not been used to analyze specific physical phenomena.
In this paper, in order to apply our model to cosmology and black holes in later sections, we give the value of $g^a{_{bc}}$ at the end of this section.

We define the covariant derivative as 
\begin{eqnarray}
\label{8.3}
{\hat h}_\mu  \equiv \partial_\mu  - i  {\hat B}_\mu 
\end{eqnarray}
using this gauge field.
Since the tetrad in teleparallel gravity is defined by applying covariant derivative to the coordinates on the tangent space (\ref{5.6}), it is similarly defined as 
\begin{eqnarray}
\label{8.4}
{\hat h}^a{_\mu} \equiv {\hat h}_\mu x^a = e^a{_\mu} + {\hat B}^a{_\mu},
\end{eqnarray}
in the noncommutative spacetime.
From (\ref{8.4}), we can see that the influence of spacetime noncommutativity also appears on the tetrad through ${\hat B}^a{_\mu}$.
Moreover, when gravity does not exist, ${\hat B}^a{_\mu}=0$, so it is obvious that ${\hat g}_{\mu\nu} = \eta_{\mu\nu}$ in this case.
Also, in order to satisfy $\delta_{\hat \epsilon} {\hat h}^a{_\mu} = 0$ as in the classical case, $\delta_{\hat \epsilon} e^a{_\mu} = - \delta_{\hat \epsilon} {\hat B}^a{_\mu}$ is necessary.
Then, since the covariant derivative is defined, the field strength can be defined as 
\begin{eqnarray}
\label{8.5}
\hat{F}_{\mu\nu} \equiv i [\hat{h}_\mu ,\hat{h}_\nu]_\star = \partial_\mu \hat{B}_\nu - \partial_\nu \hat{B}_\mu -i [ \hat{B}_\mu , \hat{B}_\nu]_\star .
\end{eqnarray}
Its transformation is
\begin{eqnarray}
\label{8.20}
\delta_{\hat{\epsilon}} \hat{F}_{\mu\nu} = i[\hat{\epsilon}, \hat{F}_{\mu\nu} ]_\star = \frac{i}{2} [\hat{\epsilon}^b, \hat{F}^c{_{\mu\nu}} ]_\star g^a{_{bc}}P_a.
\end{eqnarray}
The field strength is expanded for generators as 
\begin{eqnarray}
\hat{F}^a{_{\mu\nu}}P_a = \bigg( \partial_\mu \hat{B}^a{_\nu} - \partial_\nu \hat{B}^a{_\mu} - \frac{i}{2} [ \hat{B}^b{_\mu} , \hat{B}^c{_\nu}]_\star g^a{_{bc}} \bigg)P_a,
\end{eqnarray}
by using (\ref{8.8}).
This is used to define Lagrangian, but again we have to be careful that the product on the noncommutative spacetime is a combination of two products of the Moyal product expressing spacetime noncommutativity and inner product of the metric, as described in (\ref{7.3}) and (\ref{7.4}).
Therefore, the Lagrangian of noncommutative gravitational theory is defined as 
\begin{eqnarray}
\label{8.6}
{\cal L}_{NCG} = \frac{\hat h}{2\kappa}\bigg( \frac{1}{4} \eta_{ab}{\hat g}^{\mu\rho}{\hat g}^{\nu\sigma} +\frac{1}{2} {\hat h}_a{^\sigma}{\hat h}_b{^\nu}{\hat g}^{\mu\rho} - {\hat h}_a{^\nu}{\hat h}_b{^\sigma}{\hat g}^{\mu\rho} \bigg)\big( {\hat F}^a{_{\mu\nu}} \star {\hat F}^b{_{\rho\sigma}} \big) ,
\end{eqnarray}
by referring to the Lagrangian of teleparallel gravity (\ref{5.17}).
Here, the volume element ${\hat h} = {\rm det}( {\hat h}^a{_{\mu}} )$ is defined as the determinant of the tetrad.

Before closing this subsection, we are going to show the invariance of the action.
Since (\ref{8.6}) is invariant to subscript exchanges $(a, \mu, \nu) \leftrightarrow (b, \rho, \sigma)$, it can be expressed as 
\begin{eqnarray}
\nonumber
{\cal L}_{NCG}
&= \frac{\hat h}{2\kappa}\bigg( \frac{1}{4} \eta_{ab}{\hat g}^{\mu\rho}{\hat g}^{\nu\sigma} +\frac{1}{2} {\hat h}_a{^\sigma}{\hat h}_b{^\nu}{\hat g}^{\mu\rho} - {\hat h}_a{^\nu}{\hat h}_b{^\sigma}{\hat g}^{\mu\rho} \bigg) {\hat F}^a{_{\mu\nu}} {\hat F}^b{_{\rho\sigma}} \\
\label{8.9}
&= \frac{\hat h}{2\kappa} \bigg( \frac{1}{4} \hat{F}^a{_{\mu\nu}} \hat{F}_a{^{\mu\nu}} + \frac{1}{2} \hat{F}^a{_{\mu\nu}} \hat{F}{^{\nu\mu}}{_a} - \hat{F}^\nu{_{a\nu}} \hat{F}{^{\mu a}}{_\mu}  \bigg).
\end{eqnarray}
In the second expression of (\ref{8.9}), subscripts are raised and lowered by using the metric and tetrad such as $\hat{A}^\mu \hat{g}_{\mu\nu} = \hat{A}_\nu$.
This operation is possible because (\ref{7.5}) and (\ref{7.6}) is defined.
The transformation of action is
\begin{eqnarray}
\nonumber
\delta_{\hat{\epsilon}} S_{NCG} 
&= \int d^4x\ \delta_{\hat{\epsilon}}{\cal L}_{NCG} \\
\nonumber
&= \int d^4x\ \frac{\hat h}{2\kappa} \Bigg\{ \frac{1}{4} (\delta_{\hat{\epsilon}} \hat{F}^a{_{\mu\nu}}) \hat{F}_a{^{\mu\nu}} + \frac{1}{4}  \hat{F}^a{_{\mu\nu}} (\delta_{\hat{\epsilon}} \hat{F}_a{^{\mu\nu}}) \\
\nonumber
&\hspace{25mm}+ \frac{1}{2} (\delta_{\hat{\epsilon}}\hat{F}^a{_{\mu\nu}}) \hat{F}{^{\nu\mu}}{_a} + \frac{1}{2} \hat{F}^a{_{\mu\nu}} (\delta_{\hat{\epsilon}}\hat{F}{^{\nu\mu}}{_a}) \\
\label{8.10}
&\hspace{25mm}- (\delta_{\hat{\epsilon}}\hat{F}^\nu{_{a\nu}}) \hat{F}{^{\mu a}}{_\mu} - \hat{F}^\nu{_{a\nu}} (\delta_{\hat{\epsilon}}\hat{F}{^{\mu a}}{_\mu})  \Bigg\}.
\end{eqnarray}
At this time, the transformation of each field strength is
\begin{eqnarray}
\label{8.11}
&\delta_{\hat{\epsilon}} \hat{F}^a{_{\mu\nu}} = \frac{i}{2} [ \hat{\epsilon}^b, \hat{F}^c{_{\mu\nu}}]_\star g^a{_{bc}} = - \frac{1}{2}\theta^{\alpha\beta} \partial_\alpha \hat{\epsilon}^b \partial_\beta \hat{F}^c{_{\mu\nu}}  g^a{_{bc}},  \\
\label{8.12}
&\delta_{\hat{\epsilon}} \hat{F}_a{^{\mu\nu}} = \frac{i}{2} [ \hat{\epsilon}^b, \hat{F}{^{c\mu\nu}}]_\star g{_{abc}} =  - \frac{1}{2}\theta^{\alpha\beta} \partial_\alpha \hat{\epsilon}^b \partial_\beta \hat{F}{^{c \mu\nu}}  g{_{abc}},   \\ 
\label{8.13}
&\delta_{\hat{\epsilon}}\hat{F}{^{\nu\mu}}{_a} = \frac{i}{2} [ \hat{\epsilon}^b, \hat{F}{^{\nu\mu c}}]_\star g{_{abc}} = - \frac{1}{2}\theta^{\alpha\beta} \partial_\alpha \hat{\epsilon}^b \partial_\beta \hat{F}{^{ \nu\mu  c}}  g{_{abc}} , \\
\label{8.14}
&\delta_{\hat{\epsilon}}\hat{F}^\nu{_{a\nu}} = \frac{i}{2} [ \hat{\epsilon}^b, \hat{F}^{\nu c}{_{\nu}}]_\star g{_{abc}} = - \frac{1}{2}\theta^{\alpha\beta} \partial_\alpha \hat{\epsilon}^b \partial_\beta \hat{F}^{\nu c}{_{\nu}}  g{_{abc}}, \\
\label{8.15}
&\delta_{\hat{\epsilon}}\hat{F}{^{\mu a}}{_\mu} = \frac{i}{2} [ \hat{\epsilon}^b, \hat{F}^{\mu c}{_{\mu}}]_\star g^a{_{bc}} =  - \frac{1}{2}\theta^{\alpha\beta} \partial_\alpha \hat{\epsilon}^b \partial_\beta \hat{F}^{\mu c}{_{\mu}}  g^a{_{bc}}.
\end{eqnarray}
Also, since $\delta_{\hat{\epsilon}} \hat{h} = 0$ is satisfied, we will define
\begin{eqnarray}
\label{8.16}
i[\hat{\epsilon}, \hat{h}]_\star = - \theta^{\alpha\beta} \partial_\alpha \hat{\epsilon} \partial_\beta \hat{h} \equiv 0.
\end{eqnarray}
By using (\ref{8.11})--(\ref{8.16}), invariance of action $\delta_{\hat{\epsilon}} S_{NCG} = 0$ can be indicated at least up to the first order of noncommutative parameter.
The reason why we can not show the invariance of full action is that it is difficult to use cyclic symmetry (\ref{3.21}) because $\hat{g}_{\mu\nu}$, $\hat{h}^a{_\mu}$ and $\hat{h}$ are functions.

In this paper, as with classical teleparallel gravity, a Lorentz connection is set to zero, so our noncommutative gravitational theory is not Lorentz invariant.
If the Lorentz connection is revived, the partial derivative of the field strength (\ref{8.5}) is changed to a covariant derivative with respect to the Lorentz connection, and basically the form of Lagrangian (\ref{8.6}) is unchanged.
Therefore, our noncommutative gravitational theory may be quasi Lorentz invariant as well as classical teleparallel gravity.
In the Seiberg--Witten map, the gauge field is classically $A_\mu=0$ and the gauge field on the noncommutative spacetime satisfies ${\hat A}_\mu(A) = 0$; thus, such a field does not play an important role in noncommutative gravity either.

\subsection{\label{subsec3}Seiberg--Witten Map for Noncommutative Gravity}
In this subsection, the Seiberg--Witten map is applied to the noncommutative gravitational theory.
By applying this map, as explained in Section \ref{sec2}, to ${\hat B}_\mu$, it can be written as
\begin{eqnarray}
\label{9.1}
{\hat B}_\mu = B_\mu - \frac{1}{2}\theta^{\alpha\beta}B_\alpha ( \partial _\beta B_\mu + F_{\beta\mu} ) .
\end{eqnarray}
In the noncommutative gravitational theory based on teleparallel gravity, it is necessary to define the field strength and Lagrangian using the gauge field ${\hat B}^a{_\mu}$ developed for the generators.
By using the commutation and anticommutation relations of the generators (\ref{8.8}), ${\hat B}^a{_\mu}$ can be expressed as 
\begin{eqnarray}
\label{9.5}
{\hat B}^a{_\mu}= B^a{_\mu} - \frac{1}{4}\theta^{\alpha\beta} g^a{_{bc}}  B^b{_\alpha} ( \partial _\beta B^c{_\mu} + F^c{_{\beta\mu}} ).
\end{eqnarray}
Since the gauge field of noncommutative gravity can be written on the commutative spacetime, ${\hat g}_{\mu\nu}$ and ${\hat h}^a{_\mu}$ are expressed by classical quantities .
${\hat h}^a{_\mu}$ can be expressed as
\begin{eqnarray}
\label{9.6}
{\hat h}^a{_\mu} = h^a{_\mu} - \frac{1}{4}\theta^{\alpha\beta} g^a{_{bc}}  B^b{_\alpha} ( \partial _\beta B^c{_\mu} + F^c{_{\beta\mu}} ) ,
\end{eqnarray}
by substituting (\ref{9.5}) into (\ref{8.4}).
The metric is defined as the square of the tetrad as defined in (\ref{7.5}), and can therefore be written as
\begin{eqnarray}
\nonumber
{\hat g}_{\mu\nu} &= g_{\mu\nu}  - \frac{1}{4}\theta^{\alpha\beta} \eta_{ab} g^b{_{cd}}  B^c{_\alpha} \\
\label{9.7}
 &\hspace{25mm} \times \Big\{ h^a{_\mu} ( \partial _\beta B^d{_\nu} + F^d{_{\beta\nu}} ) 
                                                                                                          + h^a{_\nu} ( \partial _\beta B^d{_\mu} + F^d{_{\beta\mu}} ) \Big\} .
\end{eqnarray}
Here it is obvious that ${\hat g}_{\mu\nu} \rightarrow \eta_{\mu\nu}$ with the limit of $B^a{_\mu} \rightarrow 0$.
In other words, in the limit in which gravity can be ignored, the nontrivial noncommutative spacetime will smoothly transfer to the flat noncommutative spacetime defined only by the Moyal product.
Finally, the Seiberg--Witten map of ${\hat F}^a{_{\mu\nu}}$ is written as 
\begin{eqnarray}
\label{9.8}
{\hat F^a{_{\mu\nu}}} &=  F^a{_{\mu\nu}} + \frac{1}{2}\theta^{\alpha\beta} g^a{_{bc}}\big( F^b{_{\mu\alpha}} F^c{_{\nu\beta}} - B^b{_\alpha} \partial_\beta  F^c{_{\mu\nu}} \big) ,
\end{eqnarray}
in accordance with Section \ref{sec2} and (\ref{9.5}).
At this time, as in the case of (\ref{4.27}), since the classical transformation of (\ref{9.8}) is expressed as 
\begin{eqnarray}
\delta_{{\epsilon}} {\hat F^a{_{\mu\nu}}} = - \frac{1}{2}\theta^{\alpha\beta} g^a{_{bc}} \partial_\alpha \epsilon^b \partial_\beta F^c{_{\mu\nu}},
\end{eqnarray}
$\delta_{{\epsilon}} {\hat F^a{_{\mu\nu}}} = \delta_{{\hat \epsilon}} {\hat F^a{_{\mu\nu}}}$ is obtained by comparison with (\ref{8.20}).
As a result, Lagrangian is denoted as 
\begin{eqnarray}
\nonumber
{\cal L}_{NCG} =  \frac{\hat{h}}{2\kappa} \Bigg[ 
&\bigg( \frac{1}{4} T^\rho{_{\mu\nu}} T_\rho{^{\mu\nu}} + \frac{1}{2} T^\rho{_{\mu\nu}} T{^{\nu\mu}}{_\rho} - T^\rho{_{\mu\rho}}T{^{\nu\mu}}{_\nu} \bigg)\\
\nonumber
&+ \frac{1}{4} \theta^{\alpha\beta}g^a{_{bc}}\bigg\{ F_a{^{\mu\nu}} ( F^b{_{\mu\alpha}}F^c{_{\nu\beta}} - B^b{_\alpha}\partial_\beta F^c{_{\mu\nu}}) \\
\nonumber
&\hspace{20mm}+F_d{^{\mu\nu}}F^d{_{\mu a}} ( B^b{_\alpha}\partial_\beta B^c{_\nu} + B^b{_\alpha}F^c{_{\beta\nu}})\\
\nonumber
&\hspace{20mm}+2 F^{\nu\mu}{_a} (F^b{_{\mu\alpha}} F^c{_{\nu\beta}} - B^b{_\alpha} \partial_\beta F^c{_{\mu\nu}})\\
\nonumber
&\hspace{20mm}+ F^{\nu\mu}{_a} F^\rho{_{\mu\nu}} (B^b{_\alpha} \partial_\beta B^c{_\rho} + B^b{_\alpha} F^c{_{\beta\rho}})\\
\nonumber
&\hspace{20mm}+F^\mu{_{a\nu}} F^{\nu\rho}{_\mu} (B^b{_\alpha} \partial_\beta B^c{_\rho} + B^b{_\alpha} F^c{_{\beta\rho}})\\
\nonumber
&\hspace{20mm}- 4 F^{\sigma\mu}{_\sigma}( F^b{_{\mu\alpha}} F^c{_{a\beta}} + h_a{^\nu} B^b{_\alpha} \partial_\beta F^c{_{\mu\nu}})\\
\nonumber
&\hspace{20mm}- 2F^{\sigma\mu}{_{\sigma}} F^{\nu}{_{a\mu}} (B^b{_\alpha} \partial_\beta B^c{_\nu} -  B^b{_\alpha} F^c{_{\beta\nu}})\\
\label{9.10}
&\hspace{20mm}- 2F^{\sigma\nu}{_{\sigma}} F^{\mu}{_{a\mu}} (B^b{_\alpha} \partial_\beta B^c{_\nu} -  B^b{_\alpha} F^c{_{\beta\nu}})
 \bigg\}\Bigg],
\end{eqnarray}
where the volume elements $\hat{h}$ are left in the form on noncommutative spacetime.
By expanding $\hat{h}$, it can be expressed as ${\cal L}_{NCG} = {\cal L}_{TG} + {\cal L}_{NC}$, where ${\cal L}_{NC}$ is a term derived from spacetime noncommutativity.
At this time, since (\ref{9.10}) can be regarded as the Lagrangian of modified teleparallel gravity, our model expanded with noncommutative parameters is not Lorentz invariant.

In order to analyze the physical phenomena using (\ref{9.10}), it is necessary to know the components of $g^a{_{bc}}$.
At present, the components of the structure constant of the anti-commutation relation of the translational generator is unknown.
Therefore, in this research, we tentatively define these components in the following and use them for subsequent analysis.
In this case, the generator is Abelian and the anti-commutation relation can be treated as $ \{P_a,P_b \} /2 \sim P_a \otimes P_b$.
Therefore, the structure constant $g^a{_{bc}}$ of the anti-commutation relation of local translational symmetry can be represented as
\begin{eqnarray}
\label{9.2}
P_b \otimes  P_c \sim  \frac{1}{2}g^a{_{bc}} P_a.
\end{eqnarray}
In (\ref{9.2}), the left hand side includes two translational generators, on the other hand, the right hand side only one.
Thus, $g^a{_{bc}}/2$ must have a role to make single translational generator work as twice translations.
The twice classical local translational transformation $x \rightarrow x' \rightarrow x''$ can be represented as $\Phi(x'') = \Phi(x) + (\epsilon_1^a + \epsilon_2^a) \partial_a \Phi(x) = \Phi(x) + \epsilon_{3}^a \partial_a \Phi(x)$.
If this is applied to (\ref{9.2}), the local translational transformation can be reproduced by defining it as
\begin{eqnarray}
\label{9.4}
   \frac{1}{2}g^a{_{bc}} \equiv \cases{ 2/l  &for $a = b = c$ \\
                                           1/l  &for $a = b \neq c \hspace{3mm} {\rm or} \hspace{3mm} a = c \neq b$ \\
                                            0   &for $a \neq b \hspace{3mm}  {\rm and} \hspace{3mm}  a \neq c$ \\   },
\end{eqnarray}
where $l$ is a constant with dimensions of length, and its magnitude is defined as $l \sim \sqrt{|\theta^{\mu\nu}|}$.
Since $g^a{_{bc}} \sim 1/l$ appears always with $\theta^{\mu\nu}$ or the Moyal product, the equation does not break down at the commutative limit $1/l \rightarrow \infty$.
For the moment, we utilize (\ref{9.4}) in subsequent sections, but the results shown later will be qualitatively unchanged unless the order and sign of $g^a{_{bc}}$ are different from those in (\ref{9.4}).

\section{\label{sec5}Cosmology based on noncommutative gravitational theory}
Since noncommutative gravity is a theory that incorporates spacetime noncommutativity, which is an effect in regions of extreme gravity, it can be expected to show a different picture from that of standard cosmology in the early stage of the universe.
Assuming a noncommutative spacetime that becomes flat, homogeneous, and isotropic in the classical limit, classic metrics and tetrads are set to 
\begin{eqnarray}
\label{10.1}
g_{\mu\nu} &= {\rm diag} \big( -N^2(t), a^2(t), a^2(t), a^2(t) \big),\\
g^{\mu\nu} &= {\rm diag} \big( -N^{-2}(t), a^{-2}(t), a^{-2}(t), a^{-2}(t) \big), \\
\label{10.2}
h^a{_\mu} &= {\rm diag} \big( N(t), a(t) ,a(t) ,a(t) \big), \\
h_a{^\mu} &=  {\rm diag} \big( N^{-1}(t) , a^{-1}(t) ,a^{-1}(t), a^{-1}(t) \big) , \\
e^a{_\mu} &= {\rm diag} \big( 1, 1, 1, 1 \big) , \\
\label{10.3}
e_a{^\mu} &=  {\rm diag} \big(1, 1, 1, 1 \big) ,
\end{eqnarray}
where $a(t)$ is a scale factor.
Here $N(t)$ is a lapse function and $N(t) = 1$ after applying the variational principle.
Next, each component of ${\hat g}_{\mu\nu}$ and ${\hat h}^a{_\mu}$ is derived using (\ref{10.1}) and (\ref{10.2}).
Each component of $\theta^{\mu\nu}$ is 
\begin{eqnarray}
\label{10.4}
\theta^{01} = \theta^{02} = \theta^{03} = \Sigma ,\ \ \ \ \  \theta^{12} = \theta^{23} = \theta^{31} = \Theta  .
\end{eqnarray}
In (\ref{10.4}), we assume that the time direction is special in a four-dimensional spacetime, and express the time-space noncommutativity and spatial noncommutativity as $\Sigma $ and $ \Theta $, respectively.
There is sign ambiguity in the noncommutativity of time and space, but we will explain it when dealing with a cosmological bounce scenario in the next subsection and use definition (\ref{10.4}) for the time being.
In addition, since the Cartesian coordinate system is used in this paper, it is possible to define a  homogeneous and isotropic noncommutative spacetime such as (\ref{10.4}).
By substituting (\ref{10.1})--(\ref{10.3}) into (\ref{9.6}) and (\ref{9.7}), the metric and the tetrad on this spacetime can be represented as 
\begin{eqnarray}
{\hat g}_{00} &= - N^2 - 3\frac{\Sigma}{l}\bigg\{ {\dot N}N(a-1) + N(N-1){\dot a} \bigg\},& \\
{\hat g}_{0i} &=  \frac{1}{2}\frac{\Sigma}{l}\bigg\{ {\dot N}a(a-1) + (N-1){\dot a}a \bigg\},& \\
{\hat g}_{ij} &= a^2 + 8\frac{\Sigma}{l} \dot{a}a(a-1) &i = j  ,\\
{\hat g}_{ij} &=  2\frac{\Sigma}{l}\dot{a} a (a-1)  &i \neq j , \\
{\hat h}^0{_0} &= N + \frac{3}{2}\frac{\Sigma}{l}\bigg\{ {\dot N}(a-1)+(N-1){\dot a}  \bigg\} , \\
{\hat h}^i{_0} &=  \frac{1}{2}\frac{\Sigma}{l}\bigg\{ {\dot N}(a-1)+(N-1){\dot a}  \bigg\} , \\
{\hat h}^i{_j} &= a + 4\frac{\Sigma}{l} \dot{a}(a-1) &i = j ,\\
{\hat h}^i{_j} &= \frac{\Sigma}{l}\dot{a}  (a-1)  &i \neq j.
 \end{eqnarray}
Here, for convenience, note that both spacetime and tangent-space subscripts are represented by letters and that $ i = (1, 2, 3) $ represents spatial components.
In addition, the notation $\dot{a}(t) = da(t)/dt$ is used.
In general, since the $f(T)$ gravitational theory is Lorentz invariant, we must be careful to select the appropriate tetrad \cite{59}.
Our noncommutative gravitational theory can also be regarded as modified teleparallel gravity, but in this paper we use a Cartesian coordinate system consistently so there is no problem.

Note here that only $\Sigma$ exists in the metric and the tetrad.
We speculate that $\Theta$ does not appear because the spacetime is homogeneous and isotropic in the classical limit, whereas $\Sigma$ appears because the scale factor has time dependence.

Next, we derive a field equation using the tetrad and the metric.
The determinant of the tetrad on the noncommutative spacetime is denoted as 
\begin{eqnarray}
\nonumber
\hat{h} &= {\rm det}\ {\hat h}^a{_{\mu}} \\
\label{10.9}
  &= Na^3 + 12 \frac{\Sigma}{l} {\dot a}a^2(a-1) 
        +\frac{3}{2}\frac{\Sigma}{l} \big\{ {\dot N} (a-1) + (N-1){\dot a}  \big\} .
\end{eqnarray}
Therefore, the Lagrangian on the noncommutative spacetime, which is homogeneous and isotropic in classical limit, is shown as
\begin{eqnarray}
\nonumber
  {\cal L}_{NCG} &=  \frac{1}{2\kappa} \Bigg[6\frac{1}{N}a^3 \bigg( \frac{\dot a}{a} \bigg)^2 +\frac{\Sigma}{l}\Bigg\{  15\frac{1}{N}a^3\bigg( \frac{\dot a}{a} \bigg)^3 a \\
  \label{10.10}
&\hspace{20mm}+  9\frac{1}{N^2}a^3\bigg( \frac{\dot a}{a} \bigg)^3 a + 15 \frac{\dot N}{N^2} a^3  \bigg( \frac{\dot a}{a} \bigg)^2 (a-1)   \Bigg\} \Bigg] .
\end{eqnarray}
By varying (\ref{10.10}) with respect to $N(t)$, the field equation corresponding to the Friedmann equation on the vacuum-state noncommutative spacetime is derived as
\begin{eqnarray}
\label{10.11}
- \frac{3}{\kappa} a^3 \Bigg[ \bigg( \frac{ \dot{a} }{a} \bigg)^2 +  \frac{\Sigma}{l}\bigg\{ \frac{1}{2} \bigg( \frac{ \dot{a} }{a} \bigg)^3(21a-5) +5  \frac{ \dot{a} }{a}  \frac{ \ddot{a} }{a} (a-1) \bigg\}\Bigg]  = 0 .
\end{eqnarray}
Also, by varying (\ref{10.10}) with respect to $a(t)$, another equation is derived as 
\begin{eqnarray}
- \frac{3}{ \kappa}a^2 \Bigg[ 2 \frac{\ddot{a}}{a} + \bigg( \frac{\dot{a}}{a}\bigg)^2 + 8 \frac{\Sigma}{l} \bigg\{ \bigg( \frac{\dot{a}}{a}\bigg)^3 a + 3 \frac{\dot{a}}{a} \frac{\ddot{a}}{a} a   \bigg\} \Bigg] = 0 .
\end{eqnarray}

In order to derive the equations for the case where matter exists, we assume homogeneously distributed perfect fluids, rather than incorporating the coupling fermion and noncommutative gravity as in \cite{48}.
In this case, the classical energy-momentum tensor $M_a{^\rho}$ can be expressed as 
\begin{eqnarray}
\label{10.12}
M_a{^\rho} = {\rm diag} \ \bigg( - \frac{\rho}{N}, \frac{p}{a} , \frac{p}{a} , \frac{p}{a}  \bigg) ,
\end{eqnarray}
where $\rho$ is the energy density, $p$ is the pressure, and $u^\rho = (1,0,0,0)$. 
Here, the variation in the classical matter-field Lagrangian ${\cal L}_M$ is related to  $M_a{^\rho}$ as 
\begin{eqnarray}
\label{10.13}
\delta  {\cal L}_M   =  M_a{^\rho} h \ \delta h^a{_\rho}  .
\end{eqnarray}
Also note that the variations due to $N(t)$ are
\begin{eqnarray}
\delta  {\cal L}_M   =   \frac{\delta {\cal L}_M}{\delta h^a{_\rho}} \frac{\delta h^a{_\rho}}{\delta N} \delta N =  M_0{^0} h  \frac{\delta h^0{_0}}{\delta N} \delta N.
\end{eqnarray}
The energy-momentum tensor ${\hat M}_a{^\rho}$ on the noncommutative spacetime is defined as 
\begin{eqnarray} 
\label{10.15}
\delta  {\cal L}_{NCM} \equiv  {\hat M}_a{^\rho} {\hat h} \ \delta {\hat h}^a{_\rho} ,
\end{eqnarray}
using Lagrangian $\delta  {\cal L}_{NCM}$ and its Seiberg--Witten map is defined as
\begin{eqnarray}
\label{10.30}
{\hat M}_a{^\rho} \equiv M_a{^\rho} -\frac{1}{4}\theta^{\mu\nu}\eta_{ad}\eta^{ce}g^d{_{be}}B^b{_\mu}\partial_\nu M_c{^\rho},
\end{eqnarray}
with reference to  the Seiberg--Witten map of the matter field (\ref{4.24}).
By substituting (\ref{10.1})-(\ref{10.3}) and (\ref{10.12}) into (\ref{10.30}), it is apparent that ${\hat M}_a{^\rho}$ has components
 \begin{eqnarray}
{\hat M}_0{^0} &= - \frac{\rho}{N} -\frac{3}{2}\frac{\Sigma}{l}  \frac{\dot \rho}{N}(a-1) + \frac{3}{2}\frac{\Sigma}{l} \rho \frac{\dot N}{N^2} (a-1),& \\
{\hat M}_i{^0} &= \frac{1}{2} \frac{\Sigma}{l} \frac{\dot{\rho}}{N} (a-1) - \frac{1}{2} \frac{\Sigma}{l} \rho \frac{\dot N}{N^2}(a-1),  \\
{\hat M}_i{^j} &= \frac{1}{2} \frac{\Sigma}{l} \dot{p} \bigg(1-\frac{1}{a} \bigg) -\frac{1}{2}\frac{\Sigma}{l} p \frac{\dot a}{a} \bigg( 1- \frac{1}{a} \bigg)   &i \neq j , \\
{\hat M}_i{^j} &= \frac{p}{a} + \frac{\Sigma}{l} \dot{p} \bigg( 1- \frac{1}{a} \bigg) - \frac{\Sigma}{l} p \frac{\dot a}{a} \bigg( 1 - \frac{1}{a} \bigg) &i = j .
\end{eqnarray}
Therefore, the variation in ${\cal L}_{NCM}$ by $N(t)$ is 
\begin{eqnarray}
\nonumber
\delta {\cal L}_{NCM} &= \frac{\delta {\cal L}_{NCM}}{\delta N} \delta N \\
\nonumber
&= \frac{\delta {\cal L}_{NCM}}{\delta {\hat h}^a{_\rho}} \frac{\delta  {\hat h}^a{_\rho} }{\delta N} \delta N \\
\nonumber
&= \Bigg\{ {\hat M}_a{^\rho} {\hat h} \frac{\partial {\hat h}^a{_\rho} }{\partial N } - \frac{d}{dt} \bigg( {\hat M}_a{^\rho} {\hat h} \frac{\partial {\hat h}^a{_\rho} }{\partial \dot{N} } \bigg)    \Bigg\} \delta N \\
&= - \Bigg\{ a^3  \rho + \frac{15}{2} \frac{\Sigma}{l} \rho \frac{\dot{a}}{a} ( a-1) \Bigg\} \delta N.
\end{eqnarray}
Similarly, the variance due to $a(t)$ is 
\begin{eqnarray}
\nonumber
\delta {\cal L}_{NCM} &= \frac{\delta {\cal L}_{NCM}}{\delta a} \delta a \\
\nonumber
&= \frac{\delta {\cal L}_{NCM}}{\delta {\hat h}^a{_\rho}} \frac{\delta  {\hat h}^a{_\rho} }{\delta a} \delta a \\
\nonumber
&= \Bigg\{ {\hat M}_a{^\rho} {\hat h} \frac{\partial {\hat h}^a{_\rho} }{\partial a } - \frac{d}{dt} \bigg( {\hat M}_a{^\rho} {\hat h} \frac{\partial {\hat h}^a{_\rho} }{\partial \dot{a} } \bigg)    \Bigg\} \delta a \\
\label{10.16}
&=  \Bigg[ 3 p a^2  + 9 \frac{\Sigma}{l} \bigg\{ p {\dot a}a( a - 1) - \dot{p} a^2 (a-1)\bigg\}  \Bigg] \delta a .
\end{eqnarray}
Thus, from (\ref{10.11}) and (\ref{10.16}),  the Friedmann equation on the noncommutative spacetime on which the matter field exists is derived as \begin{eqnarray}
\nonumber
 &\bigg( \frac{ \dot{a} }{a} \bigg)^2 +  \frac{\Sigma}{l}\bigg\{ \frac{1}{2} \bigg( \frac{ \dot{a} }{a} \bigg)^3 \bigg( 21 \frac{a}{a_0} - 5 \bigg) +5  \frac{ \dot{a} }{a}  \frac{ \ddot{a} }{a} \bigg( \frac{a}{a_0} - 1 \bigg) \bigg\} \\
 \label{10.17}
 &\hspace{40mm}=  \frac{\kappa}{3}\rho +\frac{5}{2} \kappa \frac{\Sigma}{l} \rho \frac{\dot a}{a} \bigg( \frac{a}{a_0} - 1 \bigg) ,     \\
 \nonumber
 &2\frac{\ddot a}{a} + \bigg( \frac{\dot a}{a} \bigg)^2 + 8 \frac{\Sigma}{l} \Bigg\{ \bigg( \frac{\dot a}{a}\bigg)^3 \frac{a}{a_0} + 3 \frac{\dot a}{a} \frac{\ddot a}{a} \frac{a}{a_0}  
     \Bigg\}  \\
  \label{10.18}
 &\hspace{40mm}= - \kappa p +3 \kappa \frac{\Sigma}{l} \Bigg\{ {\dot p} \bigg( \frac{a}{a_0} - 1 \bigg) - p \frac{\dot a}{a} \bigg( \frac{a}{a_0}-1 \bigg)   \Bigg\}  .
 \end{eqnarray} 
Here we have normalized the scale factor as $a(t) \rightarrow a(t)/a_0$ using $a(t_0) = a_0$ at time $t_0$.
In this study, we deal with spacetime noncommutativity perturbatively, so $t_0$ should be a sufficiently large time.
Also, from (\ref{10.17}) and (\ref{10.18}), the expression corresponding to the energy-conservation law is derived as 
\begin{eqnarray}
\nonumber
&{\dot \rho}  - \frac{3}{2} \frac{\Sigma}{l} {\dot p} \frac{\dot a}{a} \bigg( \frac{a}{a_0} - a \bigg)   \\
\nonumber
& \hspace{7mm} = - 3 \frac{\dot a}{a} (\rho + p) - \frac{3}{2} \frac{\Sigma}{l} \Bigg\{ 5 (\rho + p) \frac{\ddot a}{a}\bigg( \frac{a}{a_0} - 1 \bigg) \\
\label{10.19}
& \hspace{50mm} + 5  \rho \bigg( \frac{\dot a}{a} \bigg)^2 +  p \bigg( \frac{\dot a}{a} \bigg)^2 \bigg( 26\frac{a}{a_0} - 11 \bigg) \Bigg\}.
\end{eqnarray}
Now, the standard matter does not satisfy the usual conservation law due to the effect of spacetime noncommutativity.
From this point of you, it can be interpreted that the terms of ${\cal O}(\Sigma/l)$ are interaction terms of the spacetime noncommutativity and the standard matter.
However, it is just naive speculation and how to interpret terms that appear due to the spacetime noncommutativity must be studied further in detail in the future.
In previous research, Wheeler--DeWitt equation based on the noncommutative Schr\"{o}dinger equation has been used when spacetime noncommutativity is treated in cosmology \cite{66,67}, and the Friedmann equation on noncommutative spacetime has not been derived.
For this reason, we can not compare the Friedmann equation on noncommutative spacetime such as (\ref{10.17}) and (\ref{10.18}) in various models, so this problem will be a future research topic.

\subsection{\label{subsec4}Spacetime Noncommutativity in the Early Universe}
In this paper, we deal with spacetime noncommutativity as a quantum effect, and since this effect is expected to appear remarkably in the early stage of the universe, we assume a radiation-dominant cosmology.
Since the relation between energy density and pressure in the radiation-dominant universe is $p = \rho/3$, energy density classically follows $\rho = (a_0^4 \rho_0) a^{-4}$ and the scale factor evolves as $ a / a_0 = (32 \pi \rho_0 / 3)^{1/4} t^{1/2}$.
In order to derive the relation between energy density and scale factor on a noncommutative spacetime, we solve the equation obtained by substituting  
\begin{eqnarray}
\label{11.1}
\rho = B a^{-4} + \frac{\Sigma}{l} \chi(t),
\end{eqnarray} 
where $B = a_0^4 \rho_0$, into (\ref{10.19}) for $\chi(t)$.
In (\ref{11.1}), we assume that the spacetime noncommutativity appears as ${\cal O}(\Sigma/l)$.
As a result, the energy density on the noncommutative spacetime can be expressed as
\begin{eqnarray}
\label{11.12}
\rho = B a^{-4} + \frac{5}{4} \frac{\Sigma}{l} B \bigg( \frac{1}{2} A^2 a^{-6} + 5 A^{2}a_0^{-1} a^{-5} - \frac{11}{2}A^2 a_0^{-2} a^{-4} \bigg),
\end{eqnarray}
where $A= (32 \pi \rho_0/3)^{1/4}a_0$.
In this case, the integration constant is determined such that the condition $\rho(t_0)=\rho_0$ (and classically $t_0=A^{-2}a_0^2$) is satisfied.
In (\ref{11.12}), the terms relating to spacetime noncommutativity exist only for higher-order scale factors than $a^{-4}$, so it is obvious that the classical term becomes dominant with the expansion of the universe.

Next, in order to investigate the effect of spacetime noncommutativity upon the scale factors in the early universe, we adopt a method to solve for $\varphi (t)$ by substituting 
\begin{eqnarray}
\label{11.2}
a(\tau) = a_0 \tau^{1/2} + \frac{\Sigma}{l}\varphi(\tau),
\end{eqnarray}
into (\ref{10.17}).
Here for convenience, time is normalized as $\tau= t / t_0$.
As a result, the scale factor in the radiation-dominant noncommutative spacetime is
\begin{eqnarray}
\label{11.3}
\frac{a(\tau)}{a_0} = \tau^{1/2} - \frac{1}{96}\sigma \big( &165\tau^{1/2} + 23 \tau^{-1/2} + 10 \tau^{-1/2} {\rm ln} (\tau) -188 \big) ,
\end{eqnarray}
where $\sigma = \Sigma / (l t_0)$ and the integration constant is determined from the condition that $ a (t_0) = a_0 $.
From (\ref{11.3}), we can read the effect of spacetime noncommutativity on the evolution of the universe.
Since the fourth term in the bracket is a minute constant, it has little influence on the universe's evolution, and the first term follows $\tau^{1/2}$ like the classical term.
For this reason, the effect of spacetime noncommutativity follows $-{\rm ln}(\tau)/\sqrt{\tau}$ and $-1/\sqrt{\tau}$ in the early universe or $-\tau^{1/2}$ after the universe has grown sufficiently.
In other words, the time evolution of spacetime noncommutativity asymptotically approaches a finite value at $\tau \rightarrow \infty$ and diverges at $\tau \rightarrow 0$.
Therefore, it can be understood that the effect of spacetime noncommutativity appeared remarkably in the early universe.

\begin{figure}[htbp]
\begin{center}
  \includegraphics[clip,width=10cm]{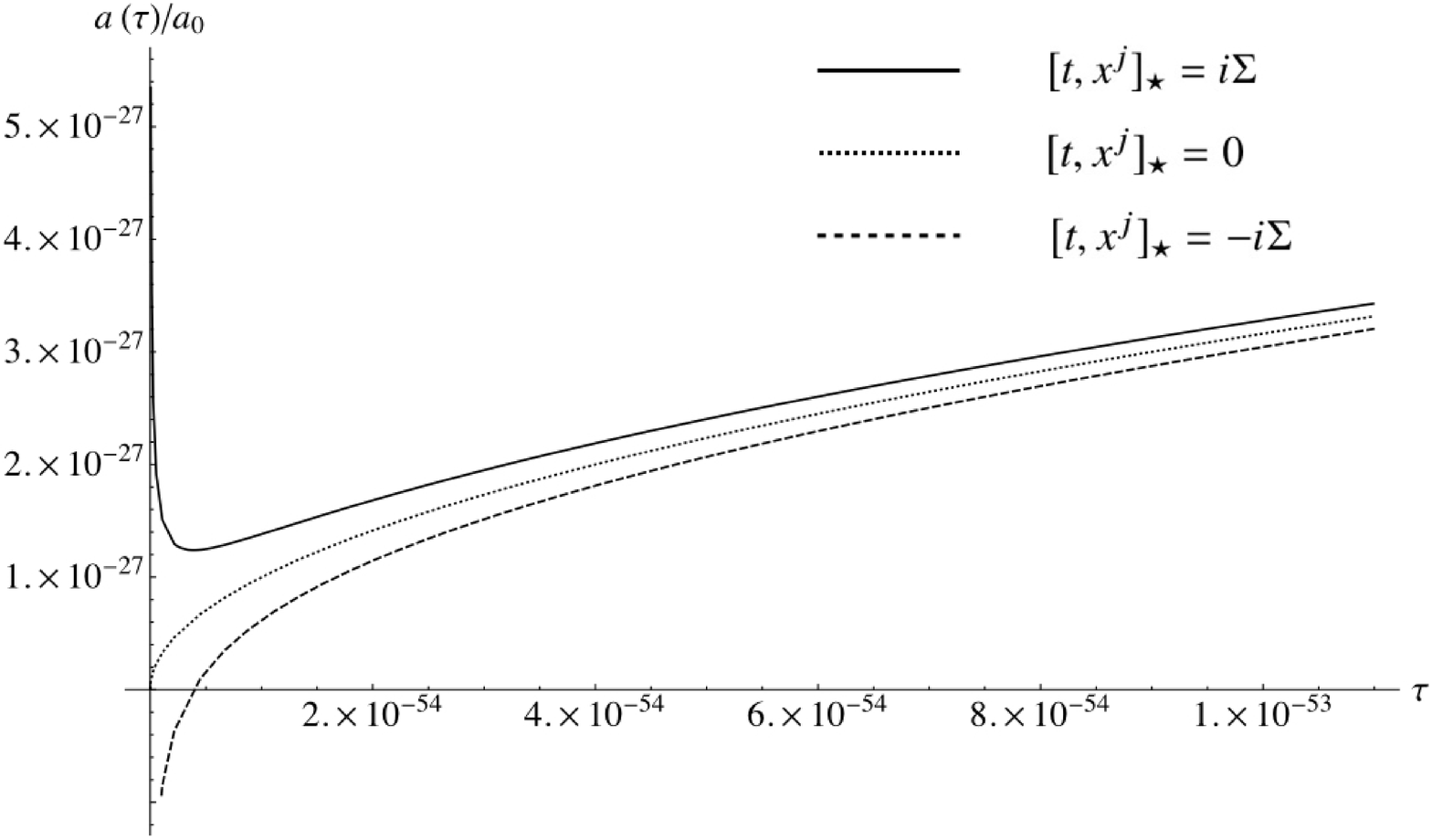}
      \caption{Time evolution of the scale factor in the early universe is plotted according to (\ref{11.3}) when the noncommutative parameter of time and space is $[t,x^j]_\star = i\Sigma,0,-i\Sigma$.}
    \label{fig2}
    \end{center}
\end{figure}

Here we reconsider the sign of the noncommutative parameter.
In (\ref{10.4}), we define the noncommutativity of time and space as $[ t, x^j ]_\star = i \Sigma$, but in fact, there is justification for whether the sign is positive or negative at that stage.
Thus, we decide the sign from the viewpoint of cosmology.
According to (\ref{11.3}), the time evolution of $a(\tau)$ in the initial universe is illustrated as shown in Fig.\ref{fig2}.
In Fig.\ref{fig2}, it is assumed that $t_0$ is around the epoch of matter-radiation equality, i.e., around $1.6 \times 10^{12} {\rm s}$, so $\sigma = 3.0 \times 10^{-56}$.
Since $a(\tau)$ expresses the size of the universe, even if quantum effects work, it is natural to require $a(\tau) \geq 0$.
From Fig.\ref{fig2}, it is obvious that the effect of a noncommutative spacetime gives a picture of the early universe different from the standard one.
In the case of $[t,x^j]_\star = - i\Sigma$ with spacetime noncommutativity, the age of the universe is somewhat smaller than that of the standard model.
However, this difference is influenced by effects in the infinitesimal region of spacetime noncommutativity, so it is negligible compared to the current age of the universe.
On the other hand, if $[t,x^j]_\star =  i\Sigma$, there is a bounce process to avoid the initial singularity of the universe evident from Fig.\ref{fig2}.
Since the bouncing process is also predicted in cosmologies based on loop quantum gravity \cite{28,32} and string theory
\cite{29}, it is natural that a similar process should also be predicted by noncommutative-gravitational cosmology.
Therefore, it can be concluded that the noncommutativity of time and space given by $[t,x^j]_\star =  i\Sigma$ is an appropriate refinement of noncommutative gravitational theory.

Because this cosmology has a bouncing process, (\ref{11.3}) has a local minimum.
The acceleration of the universe occurs from the time point $\tau_b$, satisfying $da/d\tau=0$, to the time point $\tau_c$, which satisfies $d^2a/d\tau^2=0$. 
Thereafter, the effect of spacetime noncommutativity becomes relatively small and the scale factor reduces to $a \propto \tau^{1/2}$.
By investigating the magnitude of cosmic expansion in the period $\tau_c - \tau_b$, we investigate whether spacetime noncommutativity is suitable as a source of accelerated expansion predicted by the inflation scenario \cite{31}.
First, in order to obtain the time point $\tau_b$ when $a(t)$ is a minimum value, 
\begin{eqnarray}
\label{11.4}
\frac{1}{a_0}\frac{da}{d\tau} &= \frac{1}{2}\tau^{-1/2} - \frac{1}{192}\sigma \Big( 165\tau^{-1/2} - 10\tau^{-3/2}\log (\tau) - 3 \tau^{-3/2} \Big)
\end{eqnarray}
is derived by differentiating (\ref{11.3}) with respect to $\tau$.
Therefore, the time point $\tau_b$ at which $da/d\tau = 0$ can be derived is 
\begin{eqnarray}
\label{11.5}
\tau_b = \frac{10}{3} \frac{\sigma}{32-55\sigma} W\bigg( \frac{3}{10}\frac{32-55\sigma}{\sigma} e^{-3/10} \bigg).
\end{eqnarray}
$W(x)$ is the Lambert W function, which is defined as the inverse function of $ f (x) = W e^W $ \cite{25}.
In this paper, the size of $\Sigma/l$ is set to about $\Sigma / l \sim l_{\rm Pl}$,  which is the Planck length, and even if $t_0$ is sufficiently large compared to the age of the universe, the W function in (\ref{11.5}) becomes $W \sim {\cal O}(10^2)$.
Next, differentiating (\ref{11.4}) by $\tau$ leads to 
\begin{eqnarray}
\label{11.6}
\frac{1}{a_0}\frac{d^2 a}{d \tau^2} &= -\frac{1}{4} \tau^{-3/2} + \frac{1}{384}\sigma \Big( 165\tau^{-3/2} - 30 \tau^{-5/2} \log (\tau) + 11 \tau^{-5/2}   \Big)  .
\end{eqnarray}
Therefore, the time point $\tau_c$ at which $d^2 a/d\tau^2= 0$ can be derived is
\begin{eqnarray}
\label{11.7}
\tau_c = \frac{10\sigma}{32-55\sigma} W\bigg( \frac{32-55\sigma}{10\sigma} e^{11/30}  \bigg).
\end{eqnarray}
From (\ref{11.5}) and (\ref{11.7}), it is apparent that the time period from $da/d\tau= 0$  to $d^2a/d\tau^2=0$ is 
\begin{eqnarray}
\tau_c-\tau_b \sim  \sigma \times 10^2 ,
\end{eqnarray}
when $\sigma = \Sigma/l$ is assumed to be about the Planck length.
In other words, the period required from the bottom of the bounce to the time at which the effect of spacetime noncommutativity can be ignored is around $t_c-t_b \sim t_{\rm Pl} \times 10^2$.
Furthermore, the expansion degree of the universe in this period is $a(t_c)/a(t_b) \sim {\cal O}(1)$, which clearly shows a value different from the expansion of $e^{100}$ required for inflation theory \cite{31}.
This result indicates that the bouncing process caused by spacetime noncommutativity does not offer a significant correction to the acceleration expansion in the existing inflation theory.

\section{\label{sec6}Noncommutative Schwarzschild Spacetime}
In this section, we describe the effect of spacetime noncommutativity in the vicinity of the event horizon based on a noncommutative spacetime that matches the Schwarzschild spacetime in the classical limit.
A static spherically symmetric vacuum spacetime with a point mass $M$ at its origin is classically represented as
\begin{eqnarray}
\label{12.1}
ds^2 =&  - \bigg(1-\frac{2M}{r} \bigg)dt^2 + \bigg(1-\frac{2M}{r} \bigg)^{-1} dr^2 + r^2 d\theta^2 + r^2 \sin^2 \theta d\phi ^2 
\end{eqnarray}
using a polar-coordinate system.
This is called the Schwarzschild metric, and has a metrical singularity at $r = 2M$ and a spacetime singularity at $ r = 0 $.
At metrical singularity, the component in the radial direction of the metric diverges, whereas at spacetime singularity, the component in the time direction diverges.
In particular, metrical singularity is known as the event horizon and acts as the surface of the black hole.
Here we use ``singularity" in general relativistic sense.
It should be noted that the singularity in teleparallel gravity are not yet mathematically defined and this is an unsolved problem.

The metric expressing the Schwarzschild spacetime is not limited to (\ref{12.1}).
For example, by setting
\begin{eqnarray}
\label{12.2}
r = {\tilde r} \bigg( 1 + \frac{M}{2{\tilde r}} \bigg)^2, \hspace{10mm} {\tilde r}^2 = {\tilde x}^2 + {\tilde y}^2 + {\tilde z}^2 ,
\end{eqnarray}
it can be expressed as
\begin{eqnarray}
\label{12.3}
ds^2 = - \bigg( \frac{1- M/2{\tilde r}}{1+ M/2{\tilde r}} \bigg)^2 dt^2 + \bigg( 1 + \frac{M}{2{\tilde r}} \bigg)^4 ( d{\tilde x}^2 + d{\tilde y}^2 + d{\tilde z}^2 ) ,
\end{eqnarray}
in the Cartesian coordinate system.
On the noncommutative spacetime, since it is possible to define noncommutative parameters in natural form using the Cartesian coordinate system, we will use (\ref{12.3}).
Note that by defining the radial coordinate as (\ref{12.2}), the horizon radius at (\ref{12.3}) becomes ${\tilde r} = M/2$.
In the coordinate system of (\ref{12.3}), the first term becomes zero at ${\tilde r} = M/2$ and the second term diverges at ${\tilde r} = 0 $.
In other words, by selecting the coordinate system of (\ref{12.3}), it is possible to avoid metrical singularity.
However, avoiding the metrical singularity is not equivalent to the disappearance of the event horizon.
In the following description, we will redefine the coordinates as ${\tilde r} \rightarrow r$.

Now, let us consider the noncommutative spacetime that becomes (\ref{12.3}) in the classical limit.
First, the metric on the noncommutative spacetime is derived as 
\begin{eqnarray}
\label{12.4}
{\hat g}_{00} 
&= - \bigg( \frac{1- M/2r}{1+ M/2r} \bigg)^2 - \frac{\Sigma}{l} \frac{4}{r} \bigg( \frac{M}{r} \bigg)^2  \frac{1- M/2r}{1+ M/2r}  \bigg( 1 + \frac{M}{2r} \bigg)^{-3} \frac{x+y+z}{r},  \\ 
\nonumber            
{\hat g}_{11} 
&= \bigg( 1 + \frac{M}{2r} \bigg)^4 + \frac{\Sigma}{l} \frac{2}{r}  \bigg( \frac{M}{r} \bigg)^2 \bigg(1 + \frac{11M}{8r} + \frac{3M^2}{8r^2} \bigg) \frac{x+y+z}{r}\\
\label{12.5}            
&\hspace{20mm}+ \frac{\Theta}{l} \frac{2}{r}  \bigg( \frac{M}{r} \bigg)^2 \bigg(1 + \frac{M}{4r} \bigg) \bigg( 1 + \frac{M}{2r} \bigg)^3 \frac{y-z}{r} , \\
\nonumber   
{\hat g}_{01}
& = \frac{\Sigma}{l} \frac{1}{2r} \bigg(\frac{M}{r} \bigg)^2 \frac{1 - M/2r}{1 + M/2r} \Bigg\{ \bigg( 1 + \frac{3M}{8r} + \frac{M^2}{8r^2} \bigg) \bigg( 1 + \frac{M}{2r} \bigg)^{-2} \frac{x}{r} \\
\label{12.6}
&\hspace{20mm}+ \frac{1 - M/2r}{1 + M/2r} \frac{y+z}{r}\Bigg\} - \frac{\Theta}{l} \frac{1}{r}\bigg(\frac{M}{r}\bigg)^2\bigg(1+\frac{M}{4r}\bigg)\frac{y-z}{r} ,  \\
\nonumber
{\hat g}_{12} &= - \frac{\Sigma}{l} \frac{1}{r} \bigg(\frac{M}{r}\bigg)^2 \bigg( 1 - \frac{M}{8r} - \frac{M^2}{8r^2} \bigg) \frac{x+y}{r} \\
\label{12.7}
                          &\hspace{20mm} - \frac{\Theta}{l} \frac{1}{2r} \bigg(\frac{M}{r}\bigg)^2 \bigg( 1 + \frac{M}{4r} \bigg) \bigg( 1 + \frac{M}{2r} \bigg)^3 \frac{x-y}{r},
\end{eqnarray}
based on (\ref{12.3}).
Here, other components of the metric can be easily written by cyclically interchanging the coordinates $ x $, $ y $, and $ z $ corresponding to the subscripts.
Unlike the cosmology case in the previous section, not only does $\Sigma$ show the noncommutativity of time and space, but $\Theta$ also shows spatial noncommutativity in the metric in this section.
This is presumed to be because all components of the classical metric are dependent on $r$.
Also, (\ref{12.4})$-$(\ref{12.7}) has a singularity at $r = 0$ for the noncommutative parameters.
In order to confirm that this singular point is a spacetime singularity, it is necessary to calculate the invariant over the spacetime.
However, since the physical quantity on the noncommutative spacetime in this paper is perturbed by the noncommutative parameters according to the Seiberg--Witten map, there is always a zero-order term representing the classical quantity.
Therefore, since the classical term diverges at $r=0$, it is impossible to correctly evaluate the singularities.
We expect that this will be a topic of future research on noncommutative spacetimes.

\subsection{\label{subsec5}Gravitational Redshift on a Noncommutative Spacetime}
In order to show how the effects of spacetime noncommutativity appear in the vicinity of the event horizon, we consider the gravitational redshift of light in the noncommutative spacetime.
The redshift of light means that the wavelength of light emitted from point A $\lambda_A$ and the wavelength at the observation point B $\lambda_B$ obey $ \lambda_A < \lambda_B $.
The magnitude of the light wavelength shift is expressed as 
\begin{eqnarray}
\label{13.1}
z = \frac{\lambda_B}{\lambda_A} -1 ,
\end{eqnarray}
where $z$ is the redshift parameter.
The light is redshifted when it is observed with $z > 0$ and blueshifted when it is observed with $z < 0$.
The redshift caused by gravity is obvious from the fact that infinitesimal eigentime $\Delta\tau = \sqrt{-g_{00}} \ \Delta t$ and wavelength satisfy the relation 
\begin{eqnarray}
\label{13.2}
\frac{\lambda_B}{\lambda_A} = \frac{\Delta \tau_B}{\Delta \tau_A} = \frac{\sqrt{-g_{00}(B)}}{\sqrt{-g_{00}(A)}} .
\end{eqnarray}
In other words, $z$ can be expressed as 
\begin{eqnarray}
\label{13.3}
z =  \frac{\sqrt{-g_{00}(B)}}{\sqrt{-g_{00}(A)}} -1 ,
\end{eqnarray}
when the redshift of light is caused by the influence of gravity.
Gravitational redshifts on the static vacuum spacetime can be shown by applying (\ref{12.3}) to (\ref{13.3}).
By setting the observation point $B$ at infinity, $\sqrt{-g_{00}(B)} \sim 1$ can be obtained and the redshift parameter is 
\begin{eqnarray}
\label{13.4}
z =    \frac{1+ M/2r}{1- M/2r}  -1 .
\end{eqnarray}
In this case, we choose the coordinate system (\ref{12.3}) and cannot confirm the metric singularity indicating divergence on the horizon radius $r=M/2$ in this case.
However, as is clear from (\ref {13.4}), since $ w $ diverges in $ r = M/2 $, we can see that there is an event horizon even though there is no metric singularity.

Gravitational redshifts on the noncommutative spacetime can be expressed using (\ref{12.4}), and the redshift parameter when observing the light emitted in the $x$ direction is
\begin{eqnarray}
\label{13.5}
z =& \frac{1 + M/2r}{1 - M/2r} \bigg\{ 1 + \frac{\Sigma}{l} \frac{16}{r} \bigg(\frac{M}{2r}\bigg)^2 \frac{1+M/2r}{1-M/2r}\bigg(1+\frac{M}{2r}\bigg)^{-3} \bigg\}^{-1/2} - 1 .
\end{eqnarray}
In (\ref{13.5}), the terms relating to the noncommutative parameters always take positive values because $r \geq 0$.
This means that the redshift observed in the noncommutative spacetime is weakened toward the blue side compared to the strength of the redshift in the commutative spacetime.
In other words, the quantum effect works repulsively, weakening the attractive gravity.
The redshift of the light emitted from the point where $r \gg M/2$ is expressed as
\begin{eqnarray}
\label{13.6}
z \sim \frac{M}{r} + \frac{1}{2} \bigg(\frac{M}{r} \bigg)^2 -\frac{\Sigma}{l} \frac{2}{r} \bigg(\frac{M}{r} \bigg)^2 ,
\end{eqnarray}
when terms of higher order than $(M/r)^2$ are omitted.
In (\ref{13.6}), since the third term in which $ \Sigma / l $ exists has a negative sign, it is obvious that the redshift is weakened by spacetime noncommutativity.
However, since $\Sigma/l \sim l_{ \rm Pl}$ is used when identifying spacetime noncommutativity as a quantum effect, observation of the difference from the classical value is extremely difficult using the current technology.

\begin{figure}[htbp]
\begin{center}
  \includegraphics[clip,width=9cm]{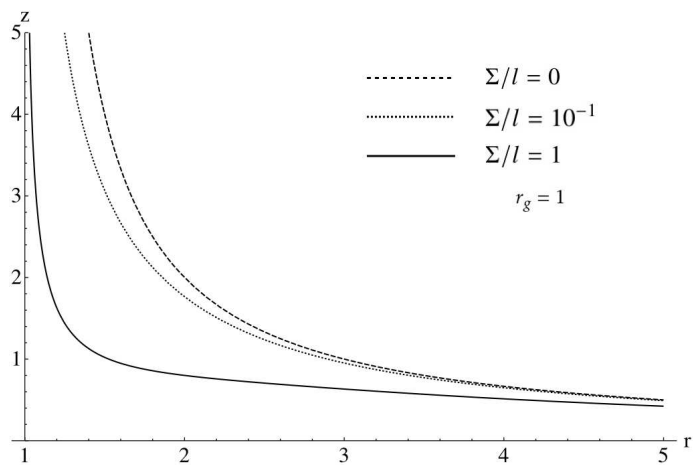}
      \caption{Plot of the gravity-redshift parameter $ z $ when $ \Sigma / l = 0, 10^{-1}, 1 $.}
    \label{fig1}
    \end{center}
\end{figure}

In Fig.\ref{fig1}, the relation between $z$ and $r$ is plotted and $r_g \equiv M/2 = 1$  and $\Sigma/l =0,10^{-1},1,$ are set for simplicity.
From each plot of Fig.\ref{fig1}, it is clear that the effect of spacetime noncommutativity weakens the gravitational redshift and its effect is more prominent at points closer to the event horizon.
Therefore, we can verify the validity of the noncommutative gravitational theory and determine the value of the noncommutative parameter by observing the light arriving from a region with extremely strong gravity near the event horizon.

Because the nature of the noncommutative spacetime weakens the gravitational redshift, we can make two guesses about the effect of spacetime noncommutativity.
The first is that the event horizon is not a perfect boundary because of spacetime noncommutativity, and the classically divergent redshift parameter becomes a finite value on a noncommutative spacetime.
Second, since the effect of the noncommutative spacetime works repulsively, the redshift of the light emitted from the vicinity of the event horizon, where the influence appears conspicuously, is weakened.
Further interpretation is possible using the cosmological result in the previous section.
In the cosmology incorporating spacetime noncommutativity, it has been shown that a bouncing process occurs, thereby indicating the existence of a repulsive force causing the expansion of the universe.
In other words, within this paper, the quantum effects on gravitation can be interpreted as a repulsive force.
Regarding black holes, it can be predicted that the results of black hole thermodynamics may change as in \cite{68} due to the influence of repulsive force in future research.

\section{\label{sec7}Conclusions}
In this paper, we developed a noncommutative gravitational theory by applying Moyal-deformation quantization and the Seiberg--Witten map to teleparallel gravity \cite{11}, a gauge theory based on local translational symmetry.
In the process, by separating the roles of metric and Moyal product, we avoided and overcame obstacles with gauge theory on nontrivial noncommutative spacetimes, such as product order and defining the summation rule for subscripts.
In this way, we could define the tetrad and the metric in a natural way, while at the same time, the physical quantity on the noncommutative spacetime was represented as a classical quantity by the Seiberg--Witten map.

Furthermore, since the noncommutative gravitational theory constructed here is based on teleparallel gravity, it is extremely simple and easy to handle, and applications to cosmology and astrophysics are easier than those in previous researches \cite{9,10,8,40,7,38,39,13,14}.
In section \ref{sec5}, we showed that there is a bounce process in which the initial singularity, which is one of the problems of classical cosmology, is avoided.
This bouncing process is also suggested by other quantum-gravity theories, such as loop-quantum gravity \cite{28,32} and string theory \cite{29}.
Section \ref{sec6} shows that in a spherically symmetric vacuum spacetime in the classical limit, the effect of spacetime noncommutativity weakens the gravitational redshift toward the blue side.
From these results, we determined that the effect of spacetime noncommutativity in gravitational theory is a repulsive force.

Note that we assumed $ | \theta^{\mu\nu} | \ll 1$ when applying the noncommutative gravitational theory to cosmology and gravitational redshift, so the results obtained here are based on a perturbed theory. 
Therefore, it is actually unknown whether phenomena such as cosmological bounce processes can be shown in non-perturbative, noncommutative gravitational theory.
Similarly, we cannot know whether $z$ diverges as Fig.\ref{fig1} in the gravitational redshift of light emitted from the vicinity of an event horizon.
However, in gravitational redshift under weak gravity, the redshift is weakened by spacetime noncommutativity as (\ref{13.6}).
In addition, it is clear that the cosmology shows a tendency for bouncing processes to exist as shown in Fig.\ref{fig2}.
Note that even though we have used a perturbed approach to applications to cosmology and gravitational redshift, both suggest that spacetime noncommutativity acts as repulsive force.

In addition, recent research \cite{46,47} has shown that spacetime noncommutativity in a perturbed Minkowski spacetime induces (Anti-) de Sitter-like curvature, using a model different to ours.
Their result indicates that spacetime noncommutativity can play the role of a cosmological constant.
In our paper, spacetime noncommutativity has the effect of a repulsive force but not a cosmological constant.
Although there are differences between their model and ours, the two models are qualitatively similar in that the cosmological constant is the source of a repulsive force.
It is therefore expected that promising suggestions for the origin of inflation and the identification of dark energy can be obtained in future research of both models.

There are several points to note about the noncommutative gravitational theory constructed in this paper.
We assumed the value of the structure constant of the anti-commutation relation as equation (\ref{9.4}), but this is not based on definite experiments or observations.
However, if the order of the values are not far from that of (\ref{9.4}) and sign is equal, the conclusions of this paper are essentially unchanged.
Also, one of the most important issue is that it is difficult to define a general coordinate transformation in noncommutative spacetime.
This issue is not limited to this approach.
Even extremely rudimentary coordinate transformations, such as transformation from Cartesian to polar coordinates, have not yet been established, and proposals have only been made in two-dimensional space \cite{26,27}.
Although methods of coordinate transformation have been proposed in previous studies based on general relativity, after all, it is only a redefinition of the noncommutative parameter in each coordinate system.
Thus, a complete formulation of noncommutative gravity has not been accomplished.
In addition, some studies define general covariance on the noncommutative spacetime using a twist that modifies the Leibniz rule, as in \cite{38,39,13,14}.
However, it is unclear whether this is physically valid.
Similarly, the noncommutative gravitational theory constructed in this paper has problems with coordinate transformation, and when transformation to the polar-coordinate system is performed naively, the metric and equation diverge at the point where the zenith angle becomes zero.
Therefore, in Sections \ref{sec5} and \ref{sec6}, the analysis was addressed using the Cartesian coordinate system consistently.
The solution to these problems would be very significant to research on noncommutative spacetimes and quantum gravity and should be examined intensively in the future.
However, in the region where quantum effects of gravity work, it is also necessary to consider the possibility that there is no theory that is invariant under a general coordinate transformation.
Naturally enough, the establishment of  an analytical method at the non-perturbative level is also an important issue.
The applicability of our model to this problem is uncertain.
However, at least our model is applicable perturbatively to cosmology and astrophysics and the results are consistent with the previous research \cite{46,47}. 
In order to develop noncommutative gravitational theory, research from various viewpoints including ours must be required.

\ack
We thank Shinta Kasuya and Takashi Kimura, who gave us invaluable feedback and warm help. 
The authors would like to thank the anonymous referees for useful comments on the manuscript.
Ryouta Matsuyama thanks Yoneda Yoshimori Education Scholarship in Kanagawa University for the grant that made it possible to complete this study.


\end{document}